\documentclass[preprint] {article}
\usepackage{amsmath,amssymb,amsfonts}
\usepackage{geometry}
\usepackage{graphicx}
\usepackage{subcaption}
\usepackage{braket}
\usepackage{float}
\usepackage{lineno}
\allowdisplaybreaks
\newcommand {\bp}{\begin{pmatrix}}
\newcommand {\ep}{\end{pmatrix}}
\newcommand{\be}{\begin{equation}} \newcommand{\ee}{\end{equation}}
\newcommand{\bea}{\begin{eqnarray}}\newcommand{\eea}{\end{eqnarray}}

\DeclareMathOperator{\sgn}{sgn}
\begin{document}
\title{On regular and chaotic dynamics of a non-${\cal{PT}}$-symmetric 
Hamiltonian system of a coupled Duffing oscillator with balanced loss and gain}

\author{ Pijush K. Ghosh\footnote{{\bf email:} 
pijushkanti.ghosh@visva-bharati.ac.in}
\ and \ Puspendu Roy\footnote {{\bf email}:puspenduroy716@gmail.com}}
\date{Department of Physics, Siksha-Bhavana, \\ 
Visva-Bharati University, \\
Santiniketan, PIN 731 235, India.}

\maketitle
\begin{abstract}

A non-${\cal{PT}}$-symmetric Hamiltonian system of a Duffing
oscillator coupled to an anti-damped oscillator with a variable
angular frequency is shown to admit periodic solutions. The result
implies that ${\cal{PT}}$-symmetry of a Hamiltonian system with
balanced loss and gain is not necessary in order to admit periodic
solutions. The Hamiltonian describes a multistable dynamical system \textemdash
three out of five equilibrium points are stable.
The dynamics of the model is investigated in detail by using
perturbative as well as numerical methods and shown to admit periodic
solutions in some regions in the space of parameters. The phase transition
from periodic to unbounded solution is to be understood without any
reference to ${\cal{PT}}$-symmetry.
The numerical analysis reveals chaotic behaviour in the system beyond
a critical value of the parameter that couples the Duffing oscillator to the
anti-damped harmonic oscillator, thereby providing the first example of
Hamiltonian chaos in a system with balanced loss and gain. The method
of multiple time-scales is used for investigating the system perturbatively. 
The dynamics of the amplitude in the leading order of the perturbation
is governed by an effective dimer model with balanced loss and gain that is
non-${\cal{PT}}$-symmetric Hamiltonian system. The dimer model is solved
exactly by using the Stokes variables and shown to admit periodic solutions
in some regions of the parameter space. 

\end{abstract}
\vspace{0.1in}
\newpage
\tableofcontents{}

\section{Introduction}

The ${\cal{PT}}$-symmetric classical Hamiltonian of coupled harmonic oscillators with
balanced loss and gain admits periodic solution
in the unbroken ${\cal{PT}}$ phase of the system\cite{ben}. The bounded
solution becomes unbounded as a result of phase-transition from an unbroken
to a broken ${\cal{PT}}$ phase. The change in the nature of solutions
accompanied by the ${\cal{PT}}$ phase-transition has important physical
consequences\cite{bpeng}. Further, the corresponding quantum system is well
defined in appropriate Stokes wedges and admits bound states in the same
unbroken ${\cal{PT}}$ phase\cite{ben}. This has lead to introduction of many
${\cal{PT}}$-symmetric Hamiltonian systems with balanced loss and gain
\cite{ben1,ivb, ds-pkg,khare, pkg-ds, ds-pkg1,p6-deb,pkg-1}. The examples
include many-particle systems\cite{ben1, pkg-ds, ds-pkg1,p6-deb}, systems with
nonlinear interaction\cite{ivb,ds-pkg, khare,pkg-ds, ds-pkg1,p6-deb},
systems with space-dependent loss-gain terms\cite{ds-pkg1}, systems with
Lorentz interaction\cite{pkg-1} etc.. The important
results which are common to all these models is that ${\cal{PT}}$-symmetric
Hamiltonian systems with balanced loss and gain may admit bounded and periodic
solutions in some regions of the space of parameters. The bounded and
unbounded solutions exist in the unbroken and broken ${\cal{PT}}$-phases of
the system, respectively. The corresponding quantum system admits bound
states in well defined Stokes wedges. A few ${\cal{PT}}$ symmetric, but
non-Hamiltonian classical systems with balanced loss and gain are also known
to share the above properties\cite{khare-0,kono}. 

The investigations on classical Hamiltonian systems with balanced loss and gain are 
mostly restricted to ${\cal{PT}}$ symmetric models. One plausible reason is that
${\cal{PT}}$-symmetric non-hermitian quantum systems admit entirely real spectra and unitary
time-evolution in unbroken ${\cal{PT}}$-phase\cite{ben-2}. The same result is valid for
a non-${\cal{PT}}$-symmetric Hamiltonian provided it is pseudo-hermitian with respect
to a positive-definite similarity operator\cite{ali} and/or admits an antilinear
symmetry. A few examples of non-${\cal{PT}}$ symmetric non-hermitian Hamiltonian admitting
entirely real spectra and unitary time evolution may be found in the Refs.\cite{ali,pkg-ph,af,
pkg-o-1,pkg-susy,pkg-many,pt-qm-0,pt-qm-1}.
It is known\cite{pkg-ds,ds-pkg1} that  systems involving a non-conventional ${\cal{T}}$-symmetry
also share the same properties. In classical mechanics, there is no notion of pseudo-hermiticity or
anti-linear symmetry and the time-reversal symmetry is unique.
Consequently, the criterion for a classical system with balanced loss and gain to
admit periodic solution is solely based on ${\cal{PT}}$-symmetry.

There are no compelling reasons to accept that only ${\cal{PT}}$-symmetric classical
Hamiltonian system with balanced loss and gain may admit periodic solutions.
There should be enough provision for accommodating non-${\cal{PT}}$-symmetric classical Hamiltonian in the
investigations on systems with balanced loss and gain, which upon quantization may lead to
a pseudo-hermitian system and/or a Hamiltonian with an anti-linear symmetry different from
the ${\cal{PT}}$-symmetry. It may be noted here that the Hamiltonian formulation of generic systems
with balanced loss and gain does not require any assumption on an underlying
discrete or continuous symmetry\cite{pkg-ds,ds-pkg1,p6-deb,pkg-1}. In particular, a Hamiltonian system in its standard
formulation is necessarily non-dissipative,
since the flow in the position-velocity state space preserves the volume. Thus, a
system for which individual degrees of freedom are subjected to loss or gain is
Hamiltonian only if the net loss-gain is zero\cite{pkg-1}. This condition for the
case of constant loss and gain essentially implies ${\cal{PT}}$ symmetry of the
term $H_0=(P_x+\gamma y)(P_y - \gamma x)$ that appears in the Hamiltonian $H=H_0+V(x,y)$,
where $P_x$ and $P_y$ are canonical conjugate momenta corresponding to the coordinates
$x$ and $y$, respectively and $\gamma$ is the gain-loss parameter\cite{ben,ben1,ivb,
ds-pkg,khare, pkg-ds, ds-pkg1,p6-deb,pkg-1}. The Hamiltonian $H_0$ is not even required
to be ${\cal{PT}}$ symmetric for the case of space-dependent loss-gain terms,
i.e. $\gamma\equiv \gamma(x,y)$\cite{p6-deb,pkg-1}. The most important point
is that the potential $V(x,y)$ need not be ${\cal{PT}}$-symmetric as far as the Hamiltonian
formulation of systems with balanced loss and gain is considered\cite{pkg-ds,ds-pkg1,p6-deb,pkg-1}. 
Further, there is no general result in a model independent way to suggest that
${\cal{PT}}$-symmetry of $V(x,y)$ and hence, of $H$ is necessary in order to have
periodic solutions. On the contrary, it is known that non-Hamiltonian 
dimer model without any ${\cal{PT}}$ symmetry due to imbalanced loss and gain admits
stable nonlinear supermodes\cite{aac}.  Similarly, within the mean field description of
Bose-Einstein condensate, stationary ground state is obtained for non-${\cal{PT}}$-symmetric confining
potential\cite{wunner}. This raises the possibility that non-${\cal{PT}}$-symmetric Hamiltonian
system with balanced loss and gain may admit periodic solutions with possible applications.
It seems that non-${\cal{PT}}$ symmetric Hamiltonian with balanced loss and
gain has not been investigated so far for system with finite degrees of freedom.

The purpose of this article is to investigate regular as well as chaotic dynamics of 
a non-${\cal{PT}}$ symmetric Hamiltonian system with balanced loss and gain.
In particular, one of the objectives is to show that transition from periodic
to unbounded solutions may be present in non-${\cal{PT}}$ symmetric Hamiltonian systems
with balanced loss and gain. The system is non-${\cal{PT}}$ symmetric to start
with and there is no question of attributing existence of these solutions to
broken or unbroken ${\cal{PT}}$-phases. Such transitions are quite common in
generic dynamical systems with or without any specific symmetry and/or an underlying
Hamiltonian structure. The inclusion of loss-gain terms with a Hamiltonian description
of the system does not make much difference. The standard techniques may 
determine the region in the parameter space in which periodic solutions are obtained.

The second objective is to investigate chaotic dynamics of a
non-${\cal{PT}}$ symmetric Hamiltonian system with balanced loss and gain. It may be
noted in this regard that chaos in ${\cal{PT}}$ symmetric systems has been studied
earlier in different contexts\cite{joshua,chaos,opto}. Classical chaos has been studied
in complex phase space for the kicked rotor and the double pendulum in Ref. \cite{joshua},
while the emphasis is on quantum kicked rotor and top in Ref. \cite{chaos}. Chaotic behaviour
in ${\cal{PT}}$-symmetric models in optomechanics and magnomechanics has also been
investigated\cite{opto}. 
No generic feature relating chaotic regime with that of broken or unbroken ${\cal{PT}}$
symmetry of the system is apparent from these investigations in a model independent way. 
Within this background, the requirement of ${\cal{PT}}$ symmetry appears to be too restrictive
to explore chaotic behaviour in a larger class of systems with balanced loss and gain which may have
possible technological applications. Thus, the emphasis is on non-${\cal{PT}}$-symmetric Hamiltonian
system with balanced loss and gain.

The Duffing oscillator is a prototype example in the study of nonlinear dynamics\cite{stro,de,de1}. It describes
a damped harmonic oscillator with an additional cubic nonlinear term. The system admits different
types of solutions in different regions of the space of parameters, including chaotic behaviour
if the system is subjected to an external forcing. A Hamiltonian formulation for the Duffing oscillator
is not known for non-vanishing damping term. However, following the methods described in Refs.
\cite{pkg-ds,p6-deb,pkg-1}, a Hamiltonian may be obtained for a system where the Duffing oscillator
is coupled to an anti-damped harmonic oscillator in a nontrivial way. The coupling to the
anti-damped oscillator effectively acts as a forcing term, thereby preparing the ground for investigating
chaotic behaviour in the system. The dynamics of the Duffing oscillator completely decouples from the
system in a particular limit, while the dynamics of the anti-damped oscillator is unidirectionally
coupled to it. It should be emphasised here that even in this limit the anti-damped oscillator is not
a time-reversed version of the Duffing oscillator. The system as a whole is non-${\cal{PT}}$-symmetric
by construction. This is the Hamiltonian system of balanced loss and gain that will be studied in detail
in this article for regular and chaotic dynamics. 

The coupled Duffing oscillator with balanced loss and gain is analyzed by perturbative as well as
numerical methods. The method of multiple scales\cite{stro,msa}
is used to find approximate solutions in the leading order of perturbation parameter. The solutions
are periodic in a region which is also obtained by linear stability analysis for the existence of
periodic solution. It should be mentioned here that the linear stability analysis may or may not be
valid for a Hamiltonian system with nonlinear interaction. However, periodic solutions are also obtained
in the same region by numerically solving the equations of motions. There are unbounded solutions outside
this region. This is one of the main results of this article \textemdash a non-${\cal{PT}}$-symmetric Hamiltonian
system with balanced loss and gain admits periodic solution. The phase transition from periodic to
unbounded solutions is specific to the model without any reference to ${\cal{PT}}$-symmetry. The
Hamiltonian is a multistable system \textemdash three out of the five equilibrium points are stable. The
phase-space of the system has a rich structure. The
bifurcation diagram of the system also shows chaotic behaviour beyond a critical value of the parameter
that couples Duffing oscillator to the anti-damped oscillator. The role of this coupling term is similar
to the forcing term of the standard forced Duffing oscillator, albeit in a nontrivial way. The chaotic
behaviour of the system is confirmed independently by various methods.
The chaotic behaviour in the coupled Duffing oscillator system with balanced loss and gain is another
important result. 

The present article also deals with a solvable non-${\cal{PT}}$-symmetric dimer Hamiltonian with balanced 
loss and gain which admits periodic solutions. The model of dimer arises as a byproduct of the
perturbative analysis of the coupled Duffing oscillator model. In particular, different choices of the
small parameters for implementing the perturbation scheme lead to different sets
of equation for the time-evolution of the amplitude which may be interpreted as models of dimers.
It is shown within this context that a model of non-${\cal{PT}}$-symmetric dimer with balanced loss
and gain is exactly solvable and admits periodic solution in some regions of the parameter space. It seems
that this is the first example of a non-${\cal{PT}}$-symmetric Hamiltonian dimer model with balanced loss and gain that admits
periodic solution.  

The plan of the article is the following. The model is introduced in Sec. 2. along with discussions
on ${\cal{PT}}$-symmetric limit of the system and linear stability analysis. A perturbative analysis
of the system by using the method of multiple-scale analysis is presented in Sec. 3.
The numerical result for the coupled Duffing-oscillator system is presented in Sec. 4.
Non-${\cal{PT}}$-symmetric Hamiltonian of a dimer with balanced loss and gain is presented in
Sec. 5. Finally, discussions on the results are presented in Sec. 6. Perturbative
analysis of the model by treating the gain-loss coefficient and the coupling constant of
the nonlinear interaction as small parameters are presented in Appendix-I in Sec. 8.

\section{The Model}

The system described by the equations of motion,
\bea
&& \ddot{x} + 2\gamma\dot{x}+\omega^{2}x+\beta_{1}y+ g x^{3}=0,\nonumber \\
&& \ddot{y} - 2\gamma\dot{y}+\omega^{2}y+\beta_{2}x+3 g x^{2}y=0,
\label{v0-duff-eqn}
\eea
\noindent is a model of coupled oscillators with nonlinear interaction and subjected
to gain and loss. The parameters $\gamma, \omega$ and $g$ correspond to the loss-gain strength,
angular frequency of the harmonic trap and the nonlinear coupling strength, respectively. The
linear coupling between the $x$ and $y$ degrees of freedom is denoted by the real constants
$\beta_1$ and $\beta_2$. The linear coupling is asymmetric for $\beta_1 \neq \beta_2$.
The above equation describes a system of balanced loss and gain in the sense that
the flow in the position-velocity state space preserves the volume, although individual
degrees of freedom are subjected to gain or loss. The system admits a Hamiltonian,
\bea
H = P_x P_y + \gamma \left ( y P_y - x P_x \right ) + \left ( \omega^2-
{\gamma^2} \right ) x y + \frac{1}{2} \left ( \beta_2 x^2 +
\beta_1 y^2 \right ) + g x^3 y,\
\label{hami}
\eea
\noindent where $P_x$ and $P_y$ are canonical momenta,
\bea
P_x=\dot{y} - \gamma y, \ \ P_y=\dot{x} + \gamma x. 
\eea
\noindent The equations of motion (\ref{v0-duff-eqn}) may be obtained from $H$.
The Hamiltonian $H$ reduces to a system of coupled harmonic oscillators with
balanced loss and gain for $g=0$ and $\beta_1=\beta_2$ which admits
periodic solutions\cite{ben} in the unbroken ${\cal{PT}}$ phase of the system.
Generalisations of the coupled oscillators model of Ref. \cite{ben} have been
considered earlier by including cubic nonlinearity and preserving the
${\cal{PT}}$-symmetry\cite{ivb,khare} of the system. As will be shown
below, the Hamiltonian $H$ for $g \neq 0$ is not ${\cal{PT}}$-symmetric
and there is no question of a broken or unbroken ${\cal{PT}}$ phases of it.
However, the system admits periodic solutions in restricted region of the
parameter space. This is a major difference from previous investigations on
models of coupled oscillators with balanced loss and gain, where the periodic
solutions are attributed to unbroken ${\cal{PT}}$-phases. 

The system described by Eqs. (\ref{v0-duff-eqn},\ref{hami}) has an interesting
limit $\beta_1=0$ for which the $x$ degree of freedom completely decouples
from the $y$ degree of freedom and describes an unforced Duffing oscillator.
The Hamiltonian $H$ for $\beta_1=0$ corresponds to the Hamiltonian of Duffing
oscillator in an ambient space of two dimensions, where the auxiliary system
is described in terms of  $y$ degree of freedom and corresponds to
a forced anti-damped harmonic oscillator with time-dependent frequency.
The time-dependence of the frequency $\omega^2 + 3 g x^2$ is implicit via
its dependence on $x$. Similarly, the time-dependence of the forcing term
$\beta_2 x$  is determined by the solutions of the
unforced Duffing oscillator. This paves the way for using well known tools and
techniques associated with a Hamiltonian system to analyse Duffing oscillator
analytically. For example, methods of canonical perturbation theory, canonical
quantization, integrability, Hamiltonian chaos etc. can be used for
investigating Duffing oscillator. The present article deals with only the
dynamical behaviour of the system for $\beta_1 \neq 0$.  

The distinction between ambient and target spaces ceases to exist for
$\beta_1 \neq 0$ and $H$ constitutes a new class of system with balanced
loss and gain. The system governed by $H$ with generic values of the parameters
may be interpreted as describing a non-standard forced Duffing oscillator
with the identification of $\beta_1 y$ in the first equation of 
(\ref{v0-duff-eqn}) as the `forcing term'. The forcing term in case
of standard Duffing oscillator can be chosen. However, for the case of
non-standard Duffing oscillator it is determined in a nontrivial way from
the second equation of (\ref{v0-duff-eqn}) which is also coupled to the first
equation. Eq. (\ref{v0-duff-eqn}) can be also interpreted as two coupled
undamped, unforced Duffing oscillators with velocity as well as space mediated
coupling terms. In particular, Eq. (\ref{v0-duff-eqn}) can be rewritten as,
\bea
&& \ddot{u} + \Omega_+ u +g u^3 +
\left \{ 2 \gamma \dot{v}+ \frac{\beta_2-\beta_1}{2} v
 - \frac{g}{2} v \left ( v^2-3 u^2 \right ) \right \} =0,\nonumber \\
&& \ddot{v} + \Omega_- v + g v^3
+ \left \{ 2 \gamma \dot{u} - \frac{\beta_2-\beta_1}{2} u
 - \frac{g}{2} u \left ( u^2-3 v^2 \right ) \right \} =0, \
\Omega_{\pm} \equiv \omega^2 \pm \frac{\beta_1+\beta_2}{2}
\eea
\noindent in the rotated co-ordinate system $(u, v)$ defined by the relations,
\bea
u=\frac{x+y}{\sqrt{2}}, \ v=\frac{x-y}{\sqrt{2}}, \ \
P_u=\frac{P_x+P_y}{\sqrt{2}}, \ P_v=\frac{P_x-P_y}{\sqrt{2}}.
\eea
\noindent In general, the coefficients $\Omega_{\pm}$ of the harmonic terms
are different and becomes identical, $\Omega_+=\Omega_-=\omega^2$ for
$\beta_1=-\beta_2$. Moreover, either $\Omega_+$ or $\Omega_-$ can be
chosen to be zero for $\omega^2 = -\frac{\beta_1+\beta_2}{2}$ or
$\omega^2 = \frac{\beta_1+\beta_2}{2}$, respectively. The loss and gain
terms are hidden in the $(u, v)$ co-ordinate system and give rise to velocity mediated
coupling between the two Duffing oscillators. The space mediated coupling between the
oscillators comprise of linear as well as nonlinear terms. The linear term vanishes for
$\beta_1=\beta_2$ and in the limit of vanishing strength of the nonlinear coupling
between the Duffing oscillators, i.e. $g \rightarrow 0$,  the system describes a linear
system that has been studied earlier\cite{ben}. The Hamiltonian in the $(u,v)$
co-ordinate system has the following form:
\bea
H_u  =  \frac{1}{2} \left (P_u-\gamma v \right )^2 - 
\frac{1}{2} \left (P_v+\gamma u \right )^2 +
\frac{\Omega_+}{2} u^2 - \frac{\Omega_-}{2} v^2 
+\frac{g}{4} \left ( u^4 - v^4 \right )
 + \frac{u v}{2} \left [ \beta_2 -\beta_1 +g (u^2-v^2) \right ].
\label{u-hami}
\eea
\noindent It is expected that some of the behaviours of the standard forced
Duffing oscillator will persist for the model under investigation. The
$(u, v)$ co-ordinates are used solely for the purpose of interpreting the
model as coupled Duffing oscillators. The rest of the discussions in this article
will be based on $(x,y)$ co-ordinates.

\subsection{Scale-transformation}

The following transformations are employed,
\bea
t \rightarrow \omega^{-1} t, \ x \rightarrow {\vert \beta_2 
\vert}^{-\frac{1}{2}} x, \
y \rightarrow {\vert \beta_1 \vert}^{-\frac{1}{2}} y, \ \beta_1 \neq 0,  \beta_2 \neq 0,
\label{scale}
\eea
\noindent in order to fix the independent scales in the system. 
This allows a reduction in total number of independent parameters which is
convenient for analyzing the system. The model can be described in terms of
three independent parameters $\Gamma, \beta$ and $\alpha$ defined as,
\bea
\Gamma = \frac{\gamma}{\omega}, \beta= \frac{\sqrt{{\vert {\beta_1} \vert}
{\vert {\beta_2}\vert}}}{\omega^2}, 
\alpha= \frac{g}{{\vert \beta_2 \vert} \omega^2},
\eea
\noindent and the equations of motion have the following expressions:
\bea
&& \ddot{x} + 2 \Gamma \dot{x}+ x + \sgn(\beta_1) \ \beta y + \alpha x^{3}=0, \nonumber \\
&& \ddot{y} - 2 \Gamma \dot{y}+ y + \sgn(\beta_2) \ \beta x+ 3 \alpha x^{2}y=0.
\label{duff-eqn}
\eea 
\noindent The sign-function $\sgn(x)$ is defined for $x \neq 0$ as $\sgn(x)=\frac{x}{\vert x \vert},
\ x \in \mathbb{R}$. The limit to the linear system $g \rightarrow 0$ now corresponds to $\alpha \rightarrow 0$.
The linear coupling between the two oscillator modes with the strength $\beta$ comprises of two
distinct cases:
\begin{itemize}
\item {\bf Linear Symmetric Coupling(LSC)}: The signs of the linear coupling terms in Eq.
(\ref{duff-eqn}) are same for this case and it occurs either for (a) $\beta_1, \beta_2 > 0$ or
(b) $\beta_1, \beta_2 < 0$. It may be noted that the linear coupling terms $( \ \sgn(\beta_1) \ \beta y,
\ \sgn(\beta_2) \ \beta x )$ appearing in Eq. (\ref{duff-eqn}) reduce to $(\beta y, \beta x)$
and $(-\beta y, -\beta x)$ for the case (a) and the case (b), respectively. These two cases
correspond to positive and negative linear coupling strengths, since $\beta > 0$. It is apparent that
Eq. (\ref{duff-eqn}) for the case (a) is related to the same equation with case (b) 
via the transformation $\beta \rightarrow -\beta$. Thus, it is suffice to consider the positive
LSC only from which the results for the negative LSC may be obtained by taking $\beta \rightarrow -\beta$.
The effect of the asymmetric linear coupling between the oscillators in Eq. (\ref{v0-duff-eqn})
is encoded in the nonlinear coupling $\alpha$ through its dependence on ${\vert \beta_2 \vert}$.

\item {\bf Linear Anti-symmetric Coupling(LAC)}: A relative sign difference between the linear
coupling terms in Eq. (\ref{duff-eqn}) is termed as LAC which may be obtained either for
$\beta_1 > 0, \beta_2 < 0$ or $\beta_1 < 0, \beta_2 >0$. It can be shown that the linear model,
i.e. $\alpha=0$, for the anti-symmetric coupling does not admit any periodic solutions. Numerical
analysis for the nonlinear model in a limited region of the parameter space indicates that it may
not admit periodic and/or stable solutions. An exhaustive numerical analysis is required to ascertain
this which is beyond the scope of this article and LAC will not be pursued further
for perturbative and numerical analysis.
\end{itemize}
\noindent The scale transformation (\ref{scale}) implies,
\be
P_x \rightarrow \frac{\omega}{\sqrt{\beta_1}} \tilde{P}_x, P_y \rightarrow
\frac{\omega}{\sqrt{\beta_2}} \tilde{P}_y, H \rightarrow \beta^{-1} \tilde{H},
\ee
\noindent where $\tilde{P}_x = \dot{y} - \Gamma y, \ \tilde{P}_y = \dot{x} + \Gamma x$ and 
\bea
\tilde{H} = \tilde{P}_x \tilde{P_y} + \Gamma \left ( y \tilde{P}_y -
x \tilde{P}_x \right ) + \left ( 1 - \Gamma^2 \right ) x y + \frac{\beta}{2}
\left [ \sgn(\beta_2) \ x^2 + \sgn(\beta_1) \ y^2\right ] + \alpha x^3 y.
\label{duff-hami}
\eea
\noindent The Hamiltonian $\tilde{H}$ and the equations of motion in (\ref{duff-eqn}) will be
considered for further analysis and the results in terms of the original variables may be obtained
by inverse scale transformations. Defining generalized momenta
$\Pi_x=\tilde{P}_x + \Gamma y, \Pi_y=\tilde{P}_y - \Gamma x$, the Hamiltonian
can be rewritten as,
\bea
\tilde{H}=\Pi_x \Pi_y + V(x,y), \ \
V(x,y)= xy + \frac{\beta}{2} \left [ \sgn(\beta_2) \ x^2 + \sgn(\beta_1) \ y^2 \right ] + \alpha x^3 y.
\eea
\noindent The Hamiltonian $\tilde{H}$ or equivalently the energy
$E=\dot{x} \dot{y} + V(x,y)$ is a constant of motion, but neither semi-positive definite nor bounded
from below. The energy may be bounded from below for specific orbits in the phase-space to be
determined from the equations of motion.

\subsection{${\cal{PT}}$-Symmetry}

The Hamiltonian may be interpreted as a two dimensional system with a
single  particle or a system of two particles in one dimension. The
Hamiltonian $H$ or $\tilde{H}$ is not ${\cal{PT}}$-symmetric for either of the cases.
For example, with the interpretation of $\tilde{H}$ describing a system of two particles
in one dimension, the parity(${\cal{P}}_1)$ and ${\cal{T}}$ symmetry are defined as,
\bea
&& {\cal{T}}: t \rightarrow -t, \ \tilde{P}_x \rightarrow - \tilde{P}_x, \
\tilde{P}_y \rightarrow - \tilde{P}_y \nonumber \\
&& {\cal{P}}_1: \ x \rightarrow - x, \ y \rightarrow - y, \tilde{P}_x
\rightarrow - \tilde{P}_x, \ \tilde{P}_y \rightarrow - \tilde{P}_y.
\eea
The term linear in $\Gamma$ is not invariant under ${\cal{P}}_1{\cal{T}}$ symmetry. Similarly,
the system is not invariant under ${\cal{PT}}$ symmetry even if the parity (${\cal{P}}$)
transformation in two dimensions is considered in its most general form:
\bea
{\cal{P}}: 
\begin{pmatrix}
{x}\\ {y}
\end{pmatrix} \rightarrow
\begin{pmatrix}
{x^{\prime}}\\ y^{\prime}
\end{pmatrix}
=
\begin{pmatrix}
{x \cos \theta + y \sin \theta}\\
{x \sin \theta - y \cos \theta}
\end{pmatrix}, \ 
\begin{pmatrix}
{\tilde{P}_x}\\ {\tilde{P}_y}
\end{pmatrix} \rightarrow
\begin{pmatrix}
{\tilde{P}_x^{\prime}}\\ \tilde{P}_y^{\prime}
\end{pmatrix}
=
\begin{pmatrix}
{\tilde{P}_x \cos \theta + \tilde{P}_y \sin \theta}\\
{\tilde{P}_x \sin \theta - \tilde{P}_y \cos \theta}
\end{pmatrix}
\eea
\noindent where $\theta \in (0, 2 \pi)$. The first term in $\tilde{H}$ is invariant under ${\cal{T}}$,
while it is invariant under ${\cal{P}}$ only for two distinct values of $\theta$, namely,
$\theta=\frac{\pi}{2}, \frac{3 \pi}{2}$. It may be noted that $\theta=\frac{\pi}{2}$ corresponds
to ${\cal{P}}: (x, y) \rightarrow (y, x), (\tilde{P}_x, \tilde{P}_y) \rightarrow
(\tilde{P}_y, \tilde{P}_x)$, while ${\cal{P}}: (x, y) \rightarrow (-y,-x),  
(\tilde{P}_x, \tilde{P}_y) \rightarrow (-\tilde{P}_y, -\tilde{P}_x)$ for $\theta=\frac{3 \pi}{2}$.
The second, third and the fourth terms of $\tilde{H}$ in
Eq. (\ref{duff-hami}) are invariant under ${\cal{PT}}$ symmetry for these two values of $\theta$
and LSC. However, the nonlinear coupling term breaks ${\cal{PT}}$ symmetry for $\alpha \neq 0$. It
may be noted that $H$ in Eq. (\ref{hami}) is not invariant under ${\cal{PT}}$ symmetry for
$\beta_1 \neq \beta_2$ even for vanishing nonlinear coupling, i.e. $g=0$. The scale transformation
plays an important role for showing implicit ${\cal{PT}}$ invariance of $H$ with $g=0$ and LSC.
The fourth term of $\tilde{H}$ breaks ${\cal{PT}}$ symmetry for the LAC and is related to
the result that $H$ is not ${\cal{PT}}$ symmetric for $\beta_1 \neq \beta_2$.  A non-vanishing nonlinear
interaction necessarily breaks ${\cal{PT}}$ invariance for $H$ and $\tilde{H}$ irrespective of
LSC or LAC.

\subsection{Stability Analysis}

The Hamilton's equations of motion,
\bea
&&\dot{x}= \tilde{P}_y - \Gamma x, \ \
\dot{y}= \tilde{P}_x + \Gamma y, \nonumber \\
&& \dot{\tilde{P}}_x= -\beta x+ \Gamma \tilde{P}_x + (\Gamma^2-1) y-
3 \alpha x^2 y,\nonumber \\
&& \dot{\tilde{P}}_y= -\beta y -\Gamma \tilde{P}_y + (\Gamma^2-1) x -
\alpha x^3,
\label{hami-eqm}
\eea
\noindent are equivalent to the  coupled second order equations (\ref{duff-eqn}) with positive
LSC. Results for negative LSC may be obtained by taking $\beta \rightarrow -\beta$.
The equilibrium points and their stability may be analyzed by employing
standard techniques. In particular, the equilibrium points are determined by the solutions of
the algebraic equations obtained by putting the right hand side of Eq. (\ref{hami-eqm}) equal
to zero. According to the Dirichlet theorem\cite{dirichlet}, an equilibrium point is stable
provided the second variation of the Hamiltonian is definite at that point. This is
neither a necessary condition nor the converse is true. If the application of the Dirichlet
theorem leads to inconclusive results, a linear stability analysis may be performed,
which is an approximate method. The method involves the study of time-evolution of small
fluctuations around an equilibrium point by keeping only linear terms. The quadratic and higher order 
fluctuations are neglected due to its smallness. The resulting linear system of coupled
differential equations can be solved to study the time-evolution of the fluctuations.
A detailed classification of equilibrium points based on the time-evolution of small fluctuation
may be found in any standard reference on nonlinear differential equation, including the
Refs. \cite{stro,center}. A Hamiltonian system admits either center i.e. closed orbit in
the phase-space surrounding an equilibrium point or hyperbolic point signalling
instability\cite{center}. The stable equilibrium point of a Hamiltonian system
necessarily corresponds to a center. It should be noted that a center is not asymptotically stable.

\subsubsection{Equilibrium Points and the Dirichlet Theorem}

The system admits five equilibrium points $P_0, P_1^{\pm},
P_2^{\pm}$ in the phase-space $(x,y,\tilde{P}_{x},\tilde{P}_{y})$ of the system, which are
determined by the solutions of the algebraic equations obtained by putting the right hand
side of Eq. (\ref{hami-eqm}) equal to zero. The equilibrium points are,
\be
P_0=(0,0,0,0), P_1^{\pm}=(\pm \delta_+,\pm\eta_+,\mp \Gamma \eta_+,\pm \Gamma
\delta_+),
P_2^{\pm}=(\pm \delta_-,\pm \eta_-,\mp \Gamma \eta_-,\pm \Gamma \delta_-),
\ee
\noindent where $\delta_{\pm}$ and $\eta_{\pm}$ are defined as follows: 
\bea
\delta_{\pm}=\frac{1}{\sqrt{3 \alpha}} \left [ -2  \pm \sqrt{ 1 + 3 \beta^2} 
\right]^{\frac{1}{2}}, \ \
\eta_{\pm}=-\frac{\delta_{\pm}}{3 \beta} \left [ 1 \pm \sqrt{1 + 3 \beta^2}
\right ].
\eea
\noindent The points $P_1^{\pm}$ are related to each other through the relation $P_1^{\pm}=-P_1^{\mp}$.
The projections of the points $P_1^{\pm}$ on the `$x-y$'-plane are related through a rotation by an angle $\pi$.
This is a manifestation of the fact that Eq. (\ref{duff-eqn}) remains invariant under the transformation
$x \rightarrow -x, y \rightarrow - y$ for fixed $\alpha, \beta, \Gamma$. Under the same transformation
$\tilde{P}_x \rightarrow - \tilde{P}_x, \tilde{P}_y \rightarrow - \tilde{P}_y$, and Eq. (\ref{hami-eqm}) remains
invariant under $(x,y,\tilde{P}_{x},\tilde{P}_{y}) \rightarrow (-x,-y,-\tilde{P}_{x},-\tilde{P}_{y})$.
The relation $P_2^{\pm}=-P_2^{\mp}$ may be explained in the same way.   

The equilibrium point $P_0$ exists all over the parameter space,
while all other points exist in restricted regions in the parameter space.
In particular,
\begin{itemize}
\item ${\bf \alpha >0}$: $P_0, P_1^{\pm}$ are equilibrium points for
$\beta^2 > 1$, while $P_0$ is the only equilibrium point for
$0 < \beta^2 \leq 1$. Points $P_2^{\pm}$ are non-existent, since
$\delta_{\pm}$ and $\eta_{\pm}$ are purely imaginary.
\item ${\bf \alpha < 0}$: $P_0, P_1^{\pm}, P_2^{\pm}$ are equilibrium points
for $0 < \beta^2 < 1$, while $P_0, P_2^{\pm}$ are equilibrium points for
$\beta^2 \geq 1$. 
\item ${\bf \alpha=0}$: Only $P_0$ is the equilibrium point.
\end{itemize}
\noindent The critical points of the Hamiltonian $\tilde{H}$ are also located at
$P_0, P_1^{\pm}, P_2^{\pm}$, since the equilibrium points correspond to
the solutions of the equations $\tilde{H}_Z \equiv \frac{\partial
\tilde{H}}{\partial Z}=0, Z\equiv(x,y,\tilde{P}_x,\tilde{P}_y)$. However, the
Hessian $\tilde{H}_{ZZ}$ of $H$ is not definite at these equilibrium points in
any region of the parameter-space. For example, the four eigenvalues of
$\tilde{H}_{ZZ}$ at $P_0$ are determined as,
\be
\frac{1}{2}\left [ \beta + \Gamma^2 \pm \left \{ 4 (1-\beta) +
(\beta+\Gamma^2)^2 \right \}^{\frac{1}{2}} \right ], \
\frac{1}{2}\left [ \beta - \Gamma^2 \pm \left \{ 4 (1+\beta) +
(\beta-\Gamma^2)^2 \right \}^{\frac{1}{2}} \right ].
\ee
\noindent The spectrum always consists of positive as well negative
eigenvalues for fixed $\Gamma$ and $\beta$ \textemdash \ all the four eigenvalues are
neither semi-positive definite nor negative-definite simultaneously. Thus,
the second variation of $\tilde{H}$ is not definite at $P_0$ and no conclusion
can be drawn about its stability by using the Dirichlet theorem\cite{dirichlet}. A similar
analytical treatment for the points $P_1^{\pm}$ and $P_2^{\pm}$ become
cumbersome. However, numerical investigations for some chosen values of the
parameters indicate that the eigenvalues of the Hessian of $H$ for none of the points
are definite.

\subsubsection{Linear Stability Analysis}

In absence of any definite results on the stability of equilibrium points
by the use of Dirichlet theorem, a linear stability analysis may be performed.
Considering small fluctuations around an equilibrium point $(x_0, y_0,
\tilde{P}_{x_0},\tilde{P}_{y_0})$ in the phase space as
$(x=x_0+\xi_1,y=y_0+\xi_2,
\tilde{P}_x=\tilde{P}_{x_0}+\xi_3,\tilde{P}_y= \tilde{P}_{y_0}+\xi_4)$ and
keeping only the terms linear in $\xi_i$ in Eq. (\ref{hami-eqm}),
the following equation is obtained,
\bea
\dot{\xi} = M \xi, \ \
M=\begin{pmatrix}
{-\Gamma} && {0} &&  {0} && {1}\\
{0} && {\Gamma} && {1} &&  {0}\\
-(\beta+6 \alpha x_0 y_0) && \Gamma^2-1-3 \alpha x_0^2&& \Gamma &&0\\
\Gamma^2-1-3 \alpha x_0^2 && -\beta && 0 && - \Gamma
\end{pmatrix},
\eea
\noindent where $\xi=(\xi_1, \xi_2, \xi_3, \xi_4)^T$ and $A^T$ denotes transpose
of $A$. The values of $x_0, y_0, \tilde{P}_{x_0}, \tilde{P}_{y_0}$ differ for
each equilibrium
point and may be substituted at an appropriate step of the stability analysis.
The characteristic equation of the matrix $M$ and its solutions $\pm i \lambda_j,
j=1, 2$ are determined as, 
\bea
&& \lambda^4 + 2 \left (1 + 3 \alpha x_0^2-2 \Gamma^2 \right ) \lambda^2 +
(1+3\alpha x_0^2)^2 -\beta^2 - 6 \alpha \beta x_0 y_0=0,\nonumber \\
&& \lambda_{j}= \left [ 1 - 2 \Gamma^2 + 3 \alpha x_0^2 - (-1)^{j+1} \sqrt{
\beta^2 + 4 \Gamma^4 - 4 \Gamma^2(1+3 \alpha x_0^2) + 6 \alpha \beta
x_0 y_0}\right ]^{\frac{1}{2}}.
\label{charpoly}
\eea
\noindent The stable solutions are obtained in different regions of the parameter space for which
$\lambda_j \in \mathbb{Re} \ \forall \ j$. Each equilibrium point is analyzed separately for its
stability.
\begin{itemize}
\item {\bf Point $P_0$}: The equilibrium point $P_0$ corresponds to $x_0=y_0=0$ and is stable in a region of
parameter-space defined by the conditions,
\bea
-\frac{1}{\sqrt{2}} < \Gamma < \frac{1}{\sqrt{2}}, \ \
4 \Gamma^2 \left ( 1 - \Gamma^2 \right ) < \beta^2 < 1.
\label{stab-con}
\eea
\noindent These inequalities are equivalent to the following conditions:
\bea
\frac{1}{2} < \beta^2 < 1, \ \ -\Gamma_0 < \Gamma < \Gamma_0, \
\Gamma_0 \equiv \frac{1}{\sqrt{2}} \sqrt{1-\sqrt{1-\beta^2}}.
\label{stab-alt-con}
\eea
\noindent The point $P_0$ is a stable equilibrium point for any $\alpha$
and restricted values of $\beta$ and $\Gamma$ specified by the condition
(\ref{stab-con}). The system with $\alpha=0$ corresponds to the coupled oscillator
model of Ref. \cite{ben} and appears as linear part of the coupled
Duffing oscillators models\cite{ivb,khare-0}. The stable equilibrium point
$P_0$ with the same stability condition has been found for all these models.
\item {\bf Point $P_1^{\pm}$}: It may be noted that $x_0^2=\delta_+^2$ and
$x_0 y_0=\delta_+ \eta_+$ for both the points $P_1^{\pm}$. The eigenvalues
$\lambda_j$ are same for both the points $P_1^{\pm}$ and a simplified expression may be obtained
as,
\bea
\lambda_{j}^2= - 1 - 2 \Gamma^2 + S -
\frac{(-1)^{j+1} }{\sqrt{3}}\sqrt{ 3-S^2 + 12 \Gamma^4 + 12 \Gamma^2
-S (12 \Gamma^2-2)}, \ \ \ j=1,2,
\eea
\noindent where $S=\sqrt{1+3\beta^2}$. Stable solutions for these two points
exist in the same region in parameter space:
\bea
\beta^2 >1, \ \ \Gamma^2 \leq \frac{\sqrt{2}-1}{2}.
\label{stab-con-1}
\eea
\noindent The equilibrium points $P_1^{\pm}$ exist for $\alpha > 0, \beta^2 >1$
and $\alpha < 0, 0 < \beta^2 <1$. Thus, $P_1^{\pm}$ are stable for
$\alpha >0$ and unstable for $\alpha <0$. The stable equilibrium points
$P_1^{\pm}$ are specific to the nonlinear interaction of the model. 
\item {\bf Point $P_2^{\pm}$}: It may be noted that $x_0^2=\delta_-^2$ and
$x_0 y_0=\delta_- \eta_-$ for both the points $P_2^{\pm}$. Stable solutions
for these two points do not exist anywhere in parameter space, since 
\bea
- \lambda_{j}^2= 1 + 2 \Gamma^2 + S+
\frac{(-1)^{j+1} }{\sqrt{3}}\sqrt{ 3-S^2 + 12 \Gamma^4 + 12 \Gamma^2
+ S (12 \Gamma^2-2)}, \ \ \ j=1,2,
\eea
\noindent implies that at least one eigenvalue has a non-vanishing imaginary part.
\end{itemize}
The Hamiltonian describes a multistable system \textemdash \ the points $P_0$ and
$P_1^{\pm}$ are stable for $\alpha >0$ with the condition (\ref{stab-con})
for $P_0$ and the condition (\ref{stab-con-1}) for $P_1^{\pm}$. For
$\alpha < 0$, only $P_0$ is the stable point. No multistable system within
the context of systems with balanced loss and gain has been reported earlier.  

All the stable equilibrium points are `center'\cite{stro,center}, i.e. 
all the eigenvalues of $M$ are purely imaginary.  It should be kept in
mind that the linear stability analysis may or may not hold for center and/or
a Hamiltonian system when the effect of the nonlinear interaction is
considered\cite{center}. The result is only indicative for scanning a large parameter space
in order to find bounded solution by using perturbative and/or numerical
methods. It will be seen that there are bounded solution for the model under
investigation whenever conditions (\ref{stab-con}) or (\ref{stab-con-1})
are satisfied as well as in other regions of the parameter space for which no
information can be gained from the linear stability analysis.

\section{Perturbative Solution}

Introducing  the following matrices,
\bea
X=\begin{pmatrix}
x \\ y
\end{pmatrix}, \ \
P=\begin{pmatrix}
1 && \beta\\
\beta  && 1
\end{pmatrix}, \ \
\tilde{V}(x,y) = \begin{pmatrix} x^3\\ 3 x^2 y \end{pmatrix}
\eea
\noindent and denoting the Pauli matrices as $\sigma_a, a=1, 2, 3$ with
$\sigma_3$ taken to be diagonal, Eq. (\ref{duff-eqn}) with positive LSC can be rewritten as,
\bea
\ddot{X} + 2 \Gamma \sigma_3 \dot{X} +  P X + \alpha \tilde{V}(x) =0.
\label{col-eqn}
\eea
\noindent Results for negative LSC may be obtained by taking $\beta \rightarrow -\beta$. The
system of coupled linear oscillators with balanced loss-gain
corresponds to $\alpha=0$ and is exactly solvable. The nonlinear interaction
is treated as perturbation for $\alpha \ll 1$. The standard perturbation theory
fails due to the appearance of secular terms, which are unbounded in time and lead
to divergences in the long-time behaviour of the solutions. One of the possible remedies
is to use the method of multiple time-scales\cite{msa} in which many time-variables are
introduced temporarily by multiplying the original time $t$ with different powers of $\alpha$.
In particular, the coordinates are expressed in powers of the small parameter
$\alpha$ and multiple time-scales are introduced as follows,
\bea
T_n= \alpha^n t, \ \
X = \sum_{n=0}^{\infty} \alpha^n X^{(n)}(T_0, T_1, \dots), \ \
X^{(n)}=\begin{pmatrix} x_{n}\\ y_{n} \end{pmatrix}.
\label{var-scale}
\eea
\noindent This introduces slow and fast time scales in the system. For example, $T_{n+1}$ is
always slower than $T_n$, since $\alpha \ll 1$. Using Eq. (\ref{var-scale})
in Eq. (\ref{col-eqn}) and equating the terms with the same coefficient $\alpha^n$ to zero,
the following equations up to $O(\alpha)$ are obtained as follows:
\bea
\label{zero}
&& {\cal{O}}(\alpha^0): \frac{\partial^2 X^{(0)}}{\partial T_0^2} +
2 \Gamma \sigma_3 \frac{\partial X^{(0)}}{\partial T_0} + P X^{(0)}=0,\\
&& {\cal{O}}(\alpha): \frac{\partial^2 X^{(1)}}{\partial T_0^2} +
2 \Gamma \sigma_3 \frac{\partial X^{(1)}}{\partial T_0} + P X^{(1)}
+ 2 \frac{\partial^2 X^{(0)}}{\partial T_0 \partial T_1} +
2 \Gamma \sigma_3 \frac{\partial X^{(0)}}{\partial T_1} +
\begin{pmatrix}
x_0^3\\
3 x_0^2 y_0
\end{pmatrix} =0.
\label{order1}
\eea
\noindent The unperturbed Eq.  (\ref{zero}) has the solution,
\bea
X^{(0)} = A_0 \ e^{-i {\lambda_1} T_0} 
\begin{pmatrix}
1 \\ \eta_1
\end{pmatrix}
+ B_0 \ e^{-i {\lambda_2} T_0}
\begin{pmatrix}
1 \\ \eta_2
\end{pmatrix} + c.c., \ \
\eta_j=\frac{1}{\beta} \left ( \lambda_j^2 + 2 i \Gamma \lambda_j -1 \right ),
\label{solo0}
\eea
\noindent where $A_0 \equiv A_0(T_1, T_2, \dots) $ and
$B_0 \equiv  B_0(T_1, T_2, \dots)$ are independent of $T_0$, but, depends on
slower time scales $T_1, T_2, \dots$ and c.c. denotes complex conjugate. The
expressions for the eigenvalues $\lambda_1, \lambda_2$ are given by Eq.
(\ref{charpoly}) with $x_0=0, y_0=0$, i.e. eigenvalues associated with the
stability of the point $P_0$. In order to find dependence of $A_0$ and
$ B_0$ on $T_1$, Eq. (\ref{order1}) is to be solved by eliminating secular
terms.

The ${\cal{O}}(\alpha)$ Eq. (\ref{order1}) is a linear inhomogeneous equation
and the complementary solution is obtained by replacing $(A_0, B_0) \rightarrow
(A_1, B_1)$ in Eq. (\ref{solo0}) describing $X^{(0)}$, where 
$A_1\equiv A_1(T_1, T_2, \dots)$ and $B_1\equiv B_1(T_1, T_2, \dots)$ are
independent of $T_0$.
The particular solution $Y^{(1)}$ is determined from the equation,
\bea
\frac{\partial^2 Y^{(1)}}{\partial T_0^2} + 2 \Gamma \sigma_3
\frac{\partial Y^{(1)}}{\partial T_0}  + P Y^{(1)} = B,
\label{parti-sol}
\eea
\noindent where $-B$ is the inhomogeneous part of Eq. (\ref{order1}).
With the introduction of two complex parameters $z_j = \Gamma+i \lambda_j$
and substituting the solution $X^{(0)}$, $B$ has the following expression:
\bea
- B & = & e^{-i {\lambda_1} T_0}
\begin{pmatrix}
2 z_1^* \frac{\partial A_0}{\partial T_1} + 3 A_0 \left (
{\vert A_0 \vert}^2 + 2 {\vert B_0 \vert}^2 \right )\\
- 2 \eta_1 z_1 \frac{\partial A_0}{\partial T_1} + 
3 A_0 \left \{  {\vert A_0 \vert}^2 \left (2 \eta_1 + \eta_1^* \right )+
2  {\vert B_0 \vert}^2  \left ( \eta_1 +\eta_2 + \eta_2^* \right )\right \}
\end{pmatrix} \nonumber \\
& + &  e^{-i {\lambda_2} T_0}
\begin{pmatrix}
2 z_2^* \frac{\partial B_0}{\partial T_1} + 3 B_0 \left ( 2 {\vert A_0 \vert}^2
+  {\vert B_0 \vert}^2 \right )\\
- 2 z_2 \eta_2 \frac{\partial B_0}{\partial T_1}  + 3 B_0 \left \{
2 {\vert A_0 \vert}^2 \left (\eta_1+\eta_1^*+\eta_2\right) +
{\vert B_0 \vert}^2 \left (2 \eta_2 + \eta_2^* \right ) \right \}
\end{pmatrix} \nonumber \\
& + & e^{-3 i {\lambda_1} T_0}
A_0^3 \begin{pmatrix} 1\\ 3 \eta_1 \end{pmatrix} +
e^{-3 i {\lambda_2} T_0} B_0^3
\begin{pmatrix} 1\\ 3 \eta_2 \end{pmatrix} \nonumber \\
& + & e^{- i ({\lambda_1}+2 {\lambda_2}) T_0}
3 A_0 B_0^2 \begin{pmatrix} 1\\ \eta_1+2 \eta_2 \end{pmatrix} +
e^{- i (2 {\lambda_1}+{\lambda_2}) T_0}
3 A_0^2 B_0 \begin{pmatrix} 1\\ 2 \eta_1 + \eta_2 \end{pmatrix}\nonumber \\
& + & e^{- i (2 {\lambda_1}-{\lambda_2}) T_0}
3 A_0^2 B_0^* \begin{pmatrix} 1\\ 2 \eta_1 + \eta_2^* \end{pmatrix} +
e^{- i ({\lambda_1}-2 {\lambda_2}) T_0}
3 A_0 (B_0^*)^2 \begin{pmatrix} 1\\ \eta_1 + 2 \eta_2^* \end{pmatrix}
+ c.c.
\eea
\noindent Eq. (\ref{parti-sol}) is a linear homogeneous equation and solutions
for each term in $B$ can be obtained separately. The first two terms of $B$
and their complex conjugates are secular and needs special treatment for
obtaining solutions. Using Fredholm Alternative Theorem\footnote{A system of
linear equations $O \xi_s=B_s$ admits solutions only if $V^{\dagger} B_s=0$
for all vectors $V$ satisfying the equation $O^{\dagger} V=0$, where a
$^{\dagger}$ denotes adjoint. The constant matrix $O$ for the present case
is obtained by substituting an ansatz for the particular solution
$Y^{(1)}=\xi_s e^{-i \lambda_s T_0}, s=1,2$ for a given secular term with the
coefficient $e^{-i \lambda_s T_0}$ in Eq. (\ref{parti-sol}).}, the conditions
for obtaining particular solutions corresponding to these two secular terms are,
\bea
&& \frac{\partial A_0}{\partial T_1} + 3A_0 \left ( Q_1 {\vert A_0 \vert}^2 +
2 Q_2 {\vert B_0 \vert}^2 \right ) =0,\nonumber \\
&& \frac{\partial B_0}{\partial T_1} + 3 B_0 \left ( 2 Q_3 {\vert A_0 \vert}^2 +
Q_4 {\vert B_0 \vert}^2 \right ) =0,
\label{t1-eqn}
\eea
\noindent where the complex constants $Q_i$'s are given by,
\bea
&& Q_1 = \frac{1 + \eta_1 (2 \eta_1 + \eta_1^*)}{2(z_1^*-z_1 \eta_1^2)}, \
Q_2= \frac{1+\eta_1 (\eta_1 +\eta_2 +\eta_2^*)}{2(z_1^*-z_1 \eta_1^2)},
\nonumber \\
&& Q_3=\frac{1+\eta_2 (\eta_1 +\eta_1^* +\eta_2)}{2(z_2^*-z_2 \eta_2^2)},
Q_4 = \frac{1 + \eta_2 (2 \eta_2 + \eta_2^*)}{2(z_2^*-z_2 \eta_2^2)}.
\eea
\noindent A general solution of Eq. (\ref{t1-eqn}) determines the $T_1$ dependence
of the constants $(A_0, B_0)$ which is not known for generic values of the
$Q_i$'s which depend on the $\Gamma$ and $\beta$. However, it can be shown
numerically  that the $Q_i$'s are purely imaginary numbers for values of the $\Gamma$
and $\beta$ satisfying the condition (\ref{stab-con}) of linear stability.
\begin{figure}[!ht]
\centering
\begin{subfigure}{.5\textwidth}
\centering
\includegraphics[width=0.8\linewidth]{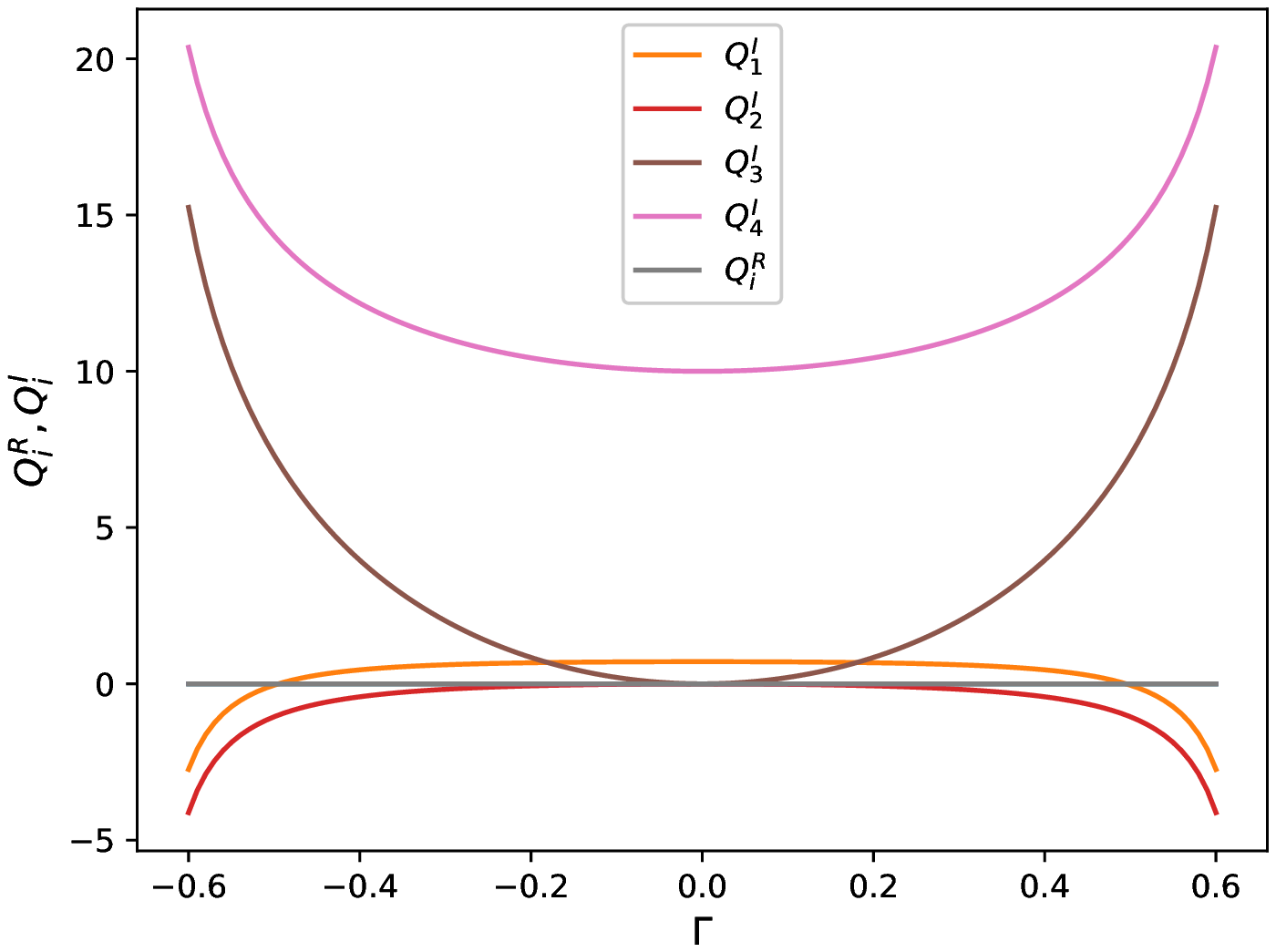}
\caption{$\beta=.99$}
\label{qplot-ga}
\end{subfigure}%
\begin{subfigure}{.5\textwidth}
\centering
\includegraphics[width=.8\linewidth]{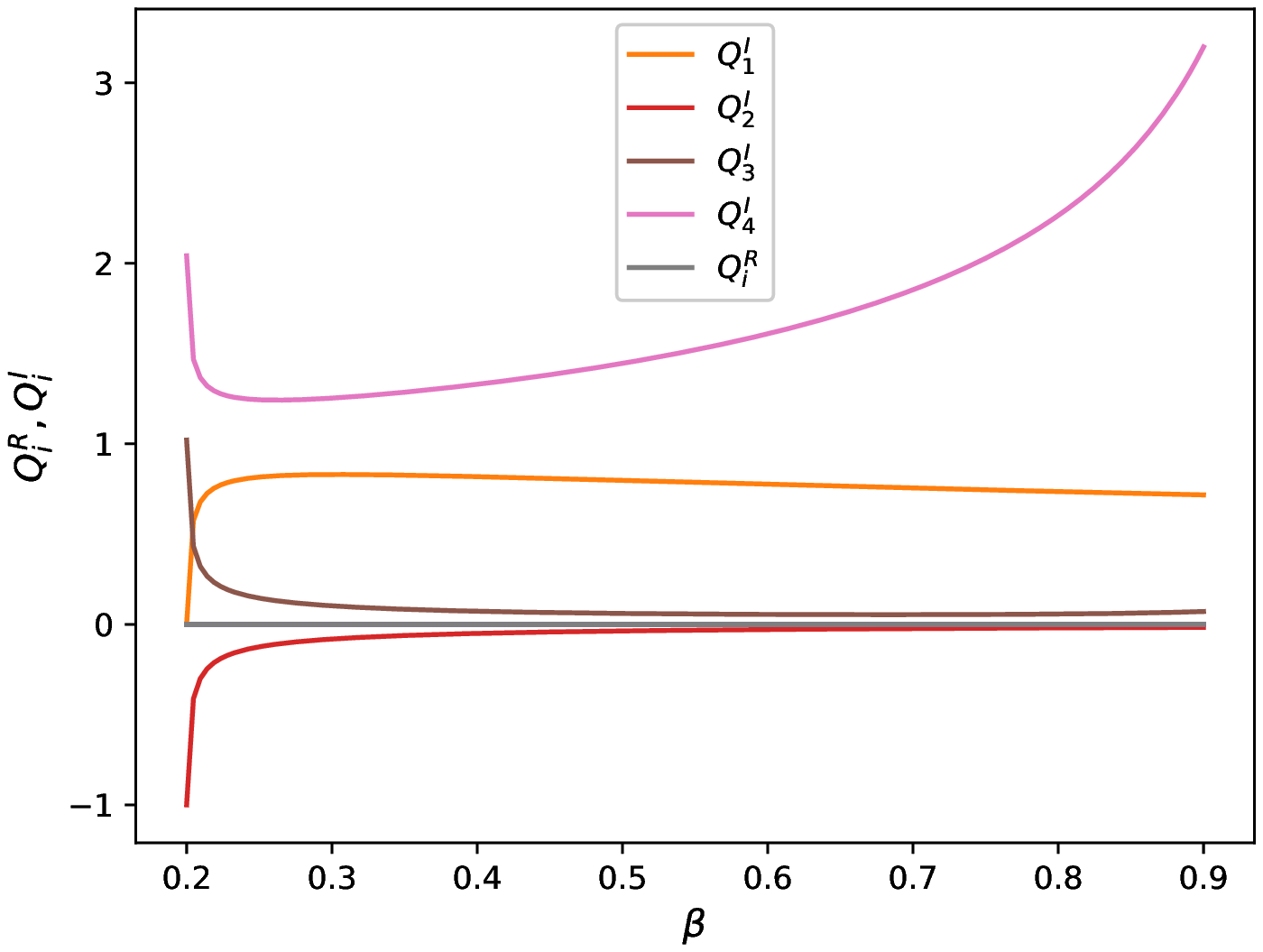}
\caption{$\Gamma=.1$}
\label{qplot-be}
\end{subfigure}%
\caption{(Color online) Plots of the real($Q_i^R$) and imaginary($Q_i^I$) parts of
the $Q_i$; (a) $Q_i^{R,I}$ versus $\Gamma$ for $\beta=.99$ and
(b) $Q_i^{R,I}$ versus $\beta$ for $\Gamma=.1$. $Q_i^R \ \forall \ i$ merge to the horizontal axis
for both the cases.}
\label{fig-qplot}
\end{figure}
It is observed numerically that the $Q_i^R \sim {\cal{O}}(10^{-16}) \ \forall \ i$
with an upper bound on the computational error of the same order, while the $Q_i^I$
take non-zero finite values for the fixed $\beta$ and $-\frac{1}{\sqrt{2}}
< \Gamma < \frac{1}{\sqrt{2}}$, where the $Q_j$ is written in terms of
real and imaginary parts as $Q_j=Q_j^R + i Q_j^I$.  Plots of the $Q_i$'s
as a function of the $\Gamma$ for fixed $\beta=.99$ is given in the Fig. (\ref{qplot-ga})
and the $Q_i$'s versus $\beta$ for $\Gamma=.1$ is given in the Fig. (\ref{qplot-be}).
It has been checked numerically for other values of $\Gamma$ and $\beta$ satisfying (\ref{stab-con})
that the same results hold.
Eq. (\ref{t1-eqn}) can be solved analytically by assuming $Q_i^R=0 \ \forall \ i$
for which ${\vert A_0 \vert}$ and ${\vert B_0 \vert}$
are constants of motion. In particular, taking the constant values
of $A_0$ and $B_0$ as their values at $t=0$, i.e. $A_0 \equiv A_0(0),
B_0 \equiv B_0(0)$, the solutions are obtained as,
\bea
&& A_0(t) = {\vert A_0(0) \vert} e^{-3 i \alpha t \left [ Q_1^I {\vert A_0(0)
\vert}^2 + 2 Q_2^I {\vert B_0(0)}^2\right ]},\nonumber \\
&& B_0 (t) = {\vert B_0(0) \vert} e^{-3 i \alpha t \left [ 2 Q_3^I
{\vert A_0(0) \vert}^2 + Q_4^I {\vert B_0(0)}^2\right ]}.
\eea
\noindent The expressions for $Q_i^I$ are not reproduced, since they are too long and does not add much
qualitative information to the discussions. The values of the constants ${\vert A_0(0) \vert}$ and
${\vert B_0(0) \vert}$ may be fixed by using initial conditions. The approximate solution is,
\bea
X & = & {\vert A_0(0) \vert} \ e^{-i t \left [ {\lambda_1} +
3 \alpha \left ( Q_1^I {\vert A_0(0) \vert}^2 + 2 Q_2^I
{\vert B_0(0)\vert}^2\right )\right ] } 
\begin{pmatrix}
1 \\ \eta_1
\end{pmatrix}\nonumber \\
& + & {\vert B_0(0) \vert} \ e^{-it \left [ {\lambda_2} +
3 \alpha \left ( 2 Q_3^I {\vert A_0(0) \vert}^2 + Q_4^I {\vert
B_0(0) \mid}^2\right ) \right ] }
\begin{pmatrix}
1 \\ \eta_2
\end{pmatrix} + c.c. + {\cal{O}}(\alpha).
\eea
\noindent The solution is bounded and consistent with linear stability
analysis. The presence of multiple time scales in the solution is apparent,
since phases and the amplitudes vary with different time scales. The solutions
are  uniform for $t \leq \alpha^{-2}$. The particular solutions of Eq.
(\ref{parti-sol}) can be obtained by
substituting $A_0(t)$ and $B_0(t)$ in $B$. However, a complete solution
which is uniform for $t \leq \alpha^{-3}$ requires to find the time dependence
of $A_1(T_1)$ and $B_1(T_1)$ appearing in the complementary solution of
$X^{(1)}$. This involves removing secular terms of differential equations
appearing at ${\cal{O}}(\alpha^2)$, which is beyond the scope of this article.

A comment is in order regarding the choice of the perturbation terms. There
are other possibilities for choosing small parameters to implement a
perturbation scheme. A few viable perturbation schemes are, (a) $\beta << 1,
\alpha << 1$, (b) $\Gamma << 1, \alpha << 1$ and (c) $\Gamma <<1, \beta << 1$.
\begin{itemize}
\item {\bf Case (a)}: A multiple scale analysis does not give any bounded
solution. The zeroth order system consists of damped and anti-damped
oscillators without any coupling between the two. There are growing as well as
decaying modes. This is consistent with the results of linear stability
analysis which predicts periodic solutions only for $\Gamma^2 < \beta^2$.
However, this condition is violated if $\beta$ is treated as small parameter,
while keeping $\Gamma$ arbitrary.
\item {\bf Case (b)}: A multiple scale analysis gives bounded solution that is
consistent with linear stability analysis. The perturbative analysis is valid
for weak nonlinear interaction characterized by $\alpha <<1$ and is described in
Appendix-I.
\item {\bf Case (c)}: The perturbation analysis is valid for strong as well as weak
nonlinearity characterized by $\alpha$. It may be noted that the discussion of this
section as well the case (b) is restricted to weak $\alpha$ only. It also deserves special
attention due to its relevance in the context of effective dimer model with balanced loss
and gain which is treated separately in Sec. 5.
\end{itemize}
All three cases correspond to perturbation around the point $P_0$. A perturbation around
the points $P_1^{\pm}$ is not pursued in  this article, since it involves use of the Jacobi
elliptic functions and the analysis of the equation governing the dynamics of the
amplitude becomes nontrivial. However, numerical solutions around the points
$P_1^{\pm}$ are provided in the next section.

\section{Numerical Solution}

The linear stability analysis of Eq. (\ref{duff-eqn}) predicts periodic
solution in regions of the parameter space defined by Eqs. (\ref{stab-con})
and (\ref{stab-con-1}). The perturbative solution obtained by using multiple
time scale analysis is also periodic in the  lowest order of the perturbation.
In absence of any global stability analysis, stability is not guaranteed for
the complete Hamiltonian including nonlinear interaction.
Further, the system may admit chaotic solution, since the first equation of
Eq. (\ref{duff-eqn}) may be interpreted as a forced Duffing
oscillator with the identification of $\beta y$ as a forcing term whose profile
is determined in a nontrivial way by the system itself. Thus, the system may
admit chaotic solution for certain regions in the parameter space. 
In this section, regular as well as chaotic dynamics of Eq. (\ref{duff-eqn})
are studied numerically. 

\begin{figure}[htbp]
\centering
\begin{subfigure}{.5\textwidth}
\centering
\includegraphics[width=0.8\linewidth]{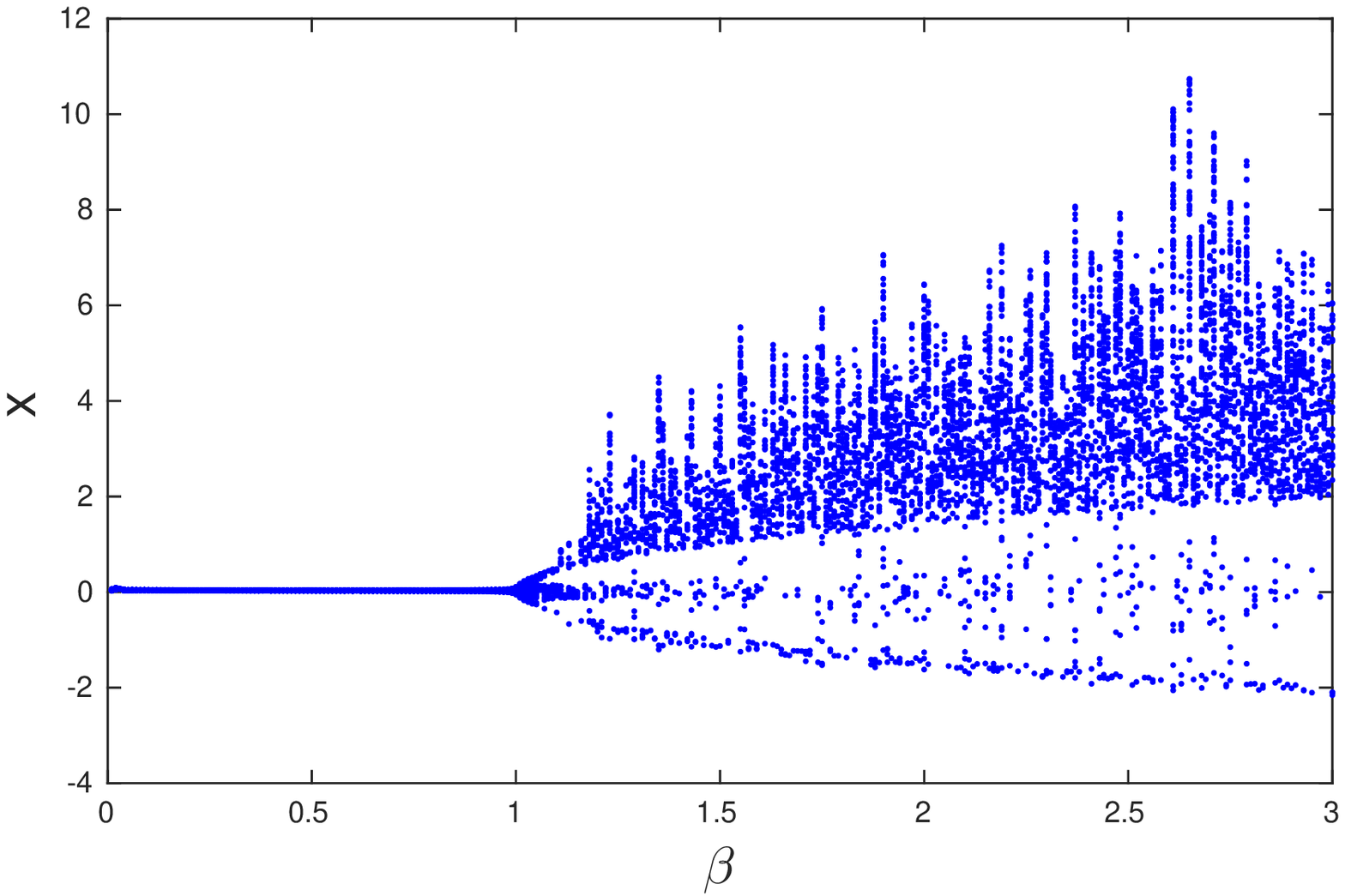}
\end{subfigure}%
\begin{subfigure}{.5\textwidth}
\centering
\includegraphics[width=.8\linewidth]{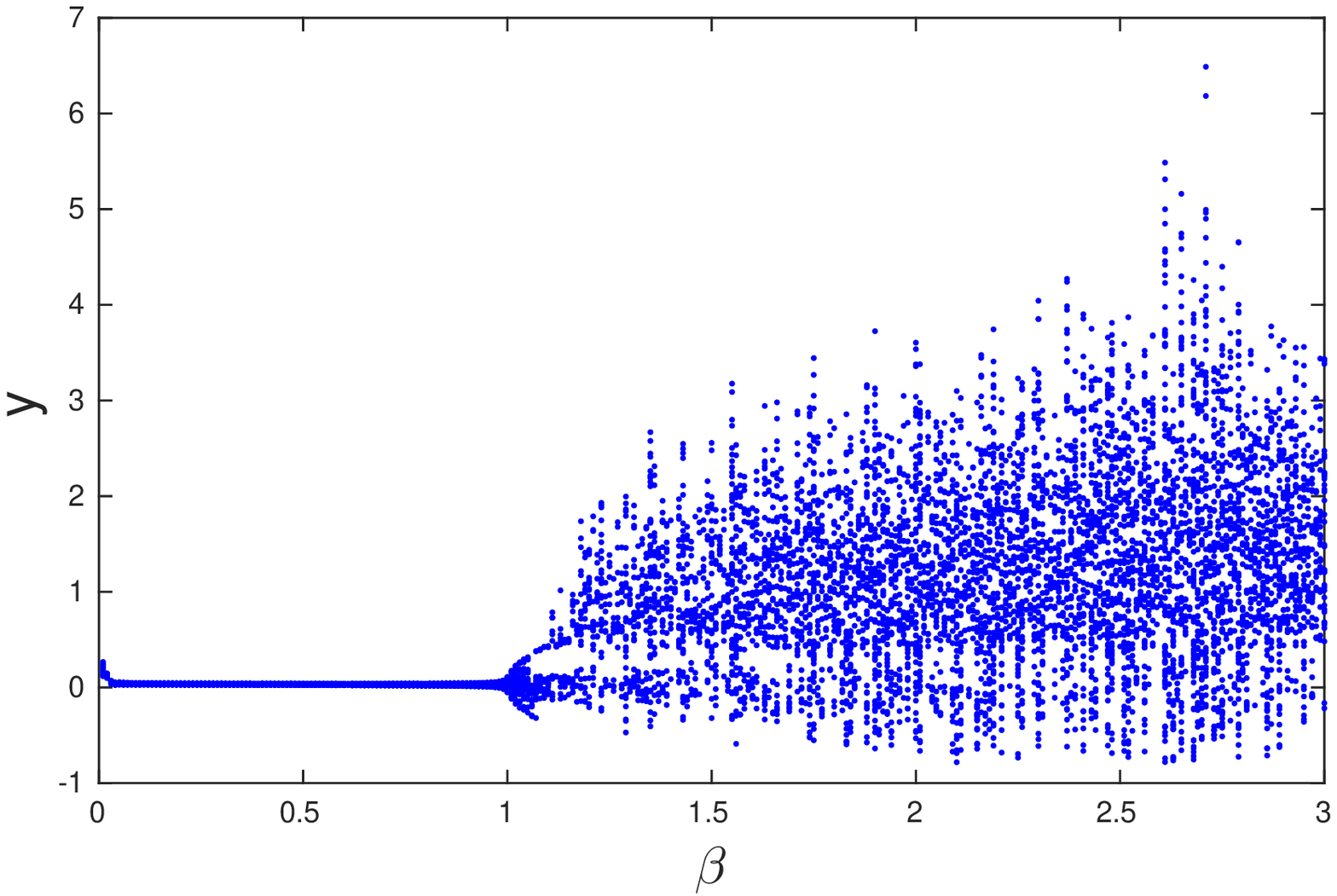}
\end{subfigure}%
\caption{(Color online) Bifurcation diagrams for $\beta$ with $\Gamma=0.01$ and $\alpha=.5 $
with the initial conditions $x(0)=0.01$, $y(0)=.02$, $\dot{x}(0)=.03$, $\dot{y}(0)=.04$}
\label{fig-bifr}
\end{figure}

The system is described in terms of three independent parameters $\Gamma, \beta$
and $\alpha$. Bifurcation diagram may be investigated by varying one of these
parameters and keeping the remaining two parameters as fixed. The bifurcation
diagram for varying $\beta$ is presented in Fig-\ref{fig-bifr} for  $\Gamma=0.01$ and $\alpha=.5 $
with the initial values of the dynamical variables near the point $P_0$.
\noindent The plot is presented for $\beta \geq 0$. However, it should be mentioned that the bifurcation
diagram is symmetric with respect to $\beta=0$, if extended to negative values of $\beta$. The onset of
chaos is seen for a critical value $\beta_c \sim 1.05 $ and persists
for $\beta \geq \beta_c$.
The crossover from regular to chaotic dynamics as $\beta$ is varied through $\beta_c$ may be understood
by interpreting the $x$ degree of freedom
as describing a forced duffing oscillator with the identification of $\beta y$ as the forcing term.
Unlike the standard forced Duffing oscillator, the forcing is determined in a nontrivial way from the
solution of the system. The chaotic behaviour of $y$ degree of freedom is induced via its coupling to
the $x$ degree of freedom. Regular and the chaotic dynamics of the system are studied in some detail in
the next two sections.  It should be mentioned here that the numerical investigations have been carried out
for very large values of $t (\sim 2,000-20, 000)$ starting from $t=0$. However, for better presentations of
the plots, figures are shown for an upper range of $t(\sim 100-2000)$ such that the qualitative features
are not lost.

\subsection{Regular Dynamics}

\begin{figure}[ht!]
\begin{subfigure}{.5\textwidth}
\centering
\includegraphics[width=.8\linewidth]{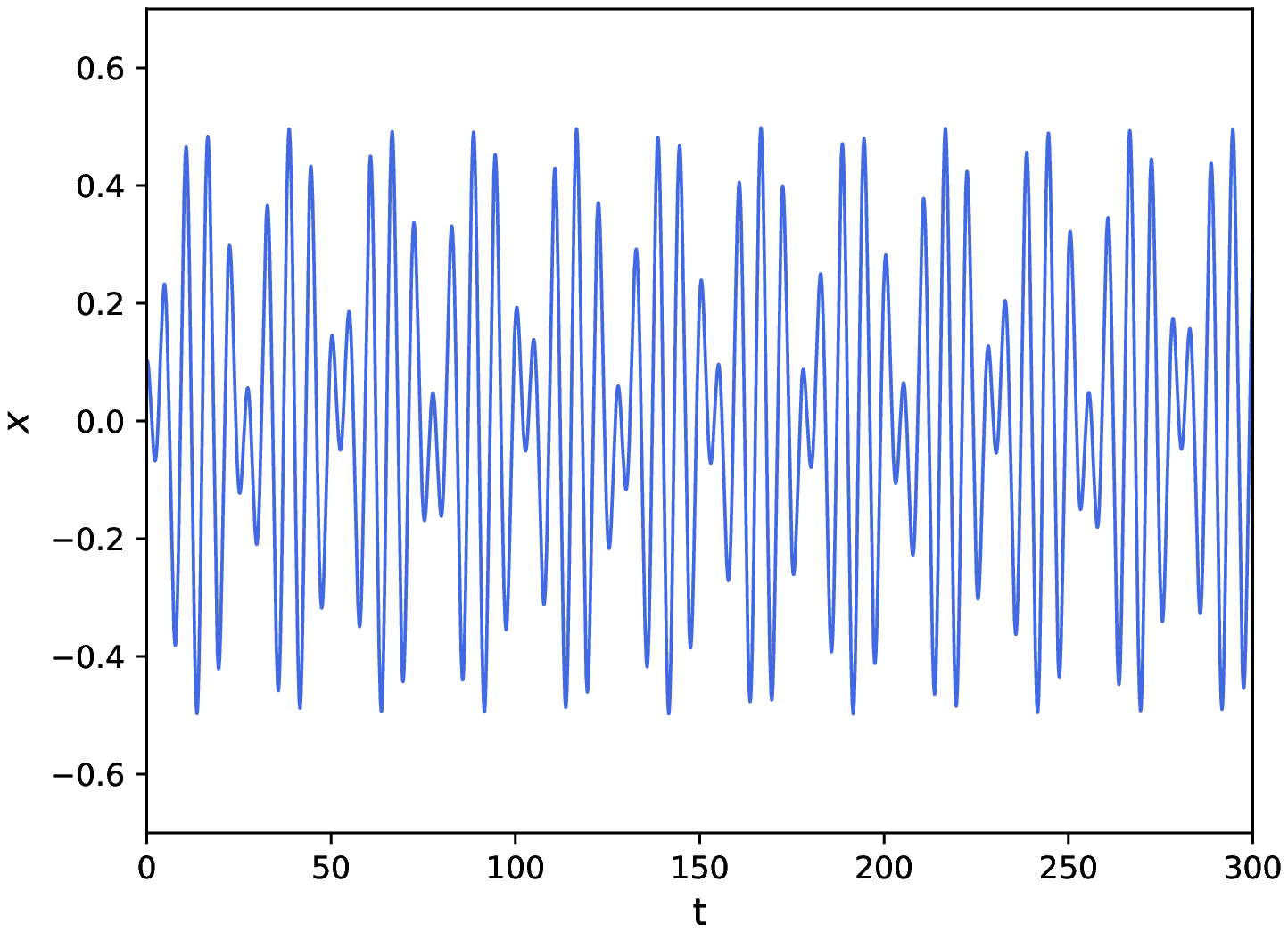} 
\caption{ $\alpha=1, \beta=.5, \Gamma=.2$}
\label{xtimeseries1}
\end{subfigure}%
\begin{subfigure}{.5\textwidth}
\centering
\includegraphics[width=.8\linewidth]{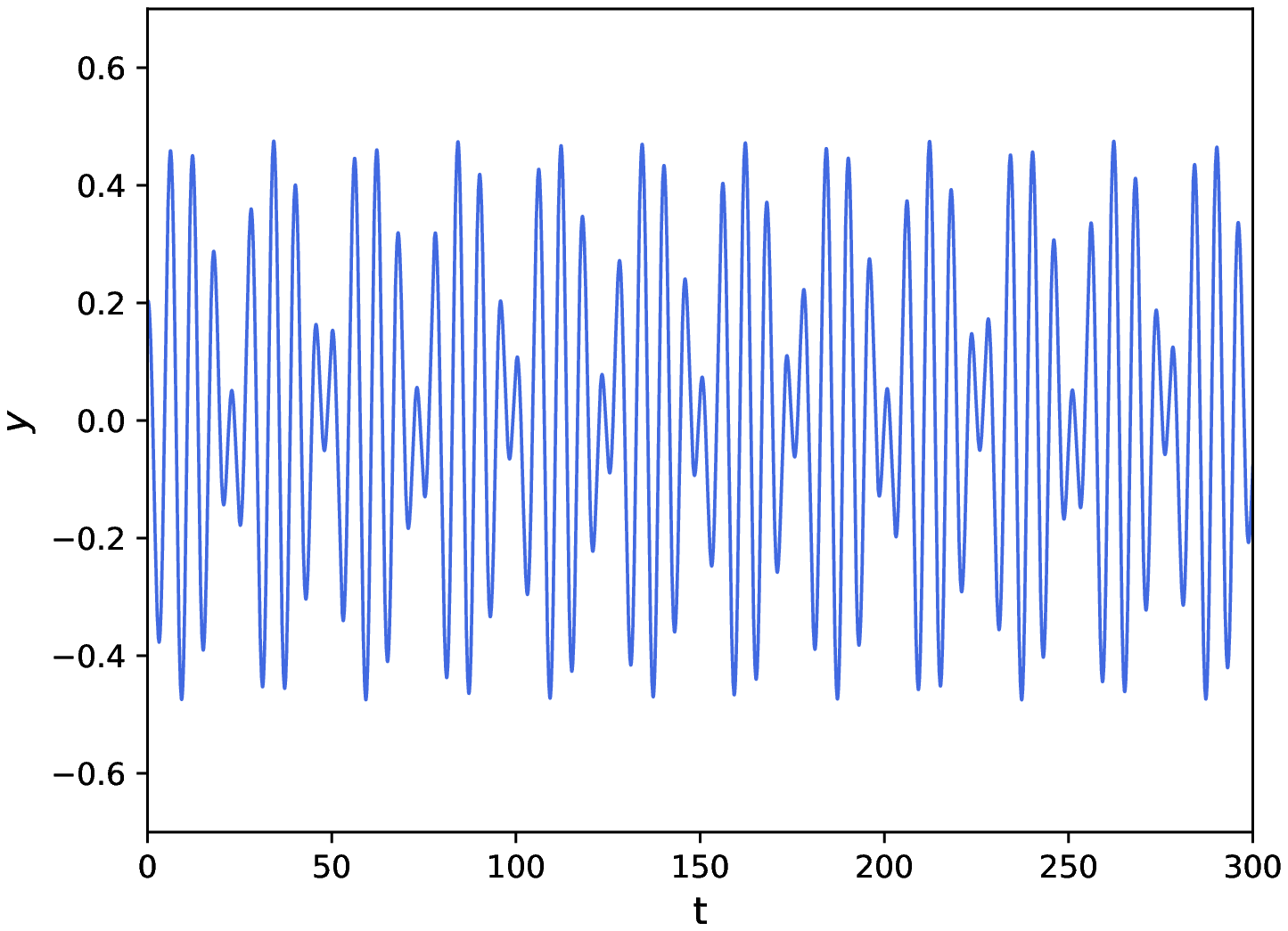}  
\caption{$\alpha=1, \beta=.5, \Gamma=.2$}
\label{ytimeseries1}
\end{subfigure}%
\newline
\begin{subfigure}{.5\textwidth}
\centering
\includegraphics[width=.8\linewidth]{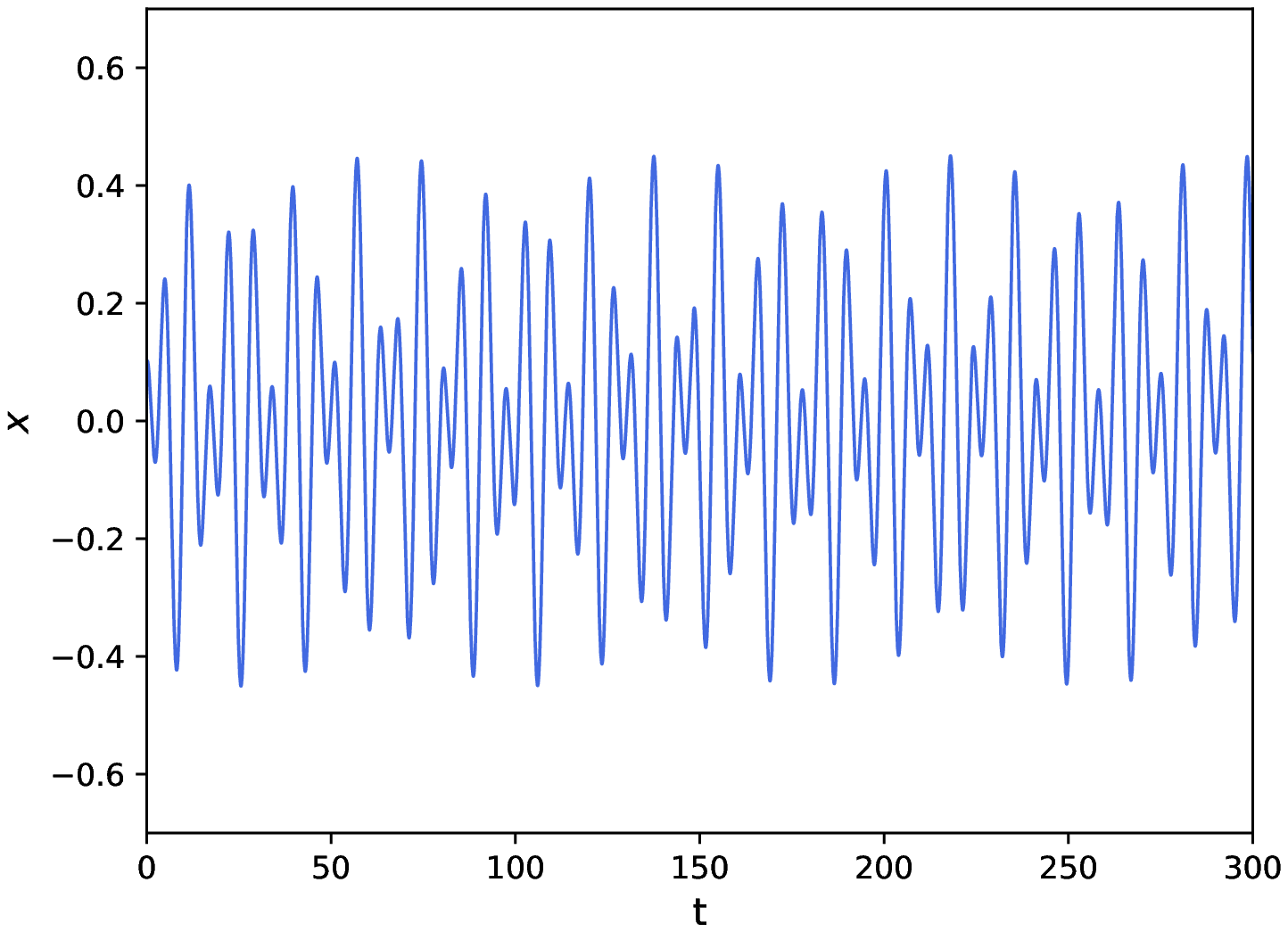}
\caption{$\alpha=-1, \beta=.5, \Gamma=.2$}
\label{xtimeseries2}
\end{subfigure}%
\begin{subfigure}{.5\textwidth}
\centering
\includegraphics[width=.8\linewidth]{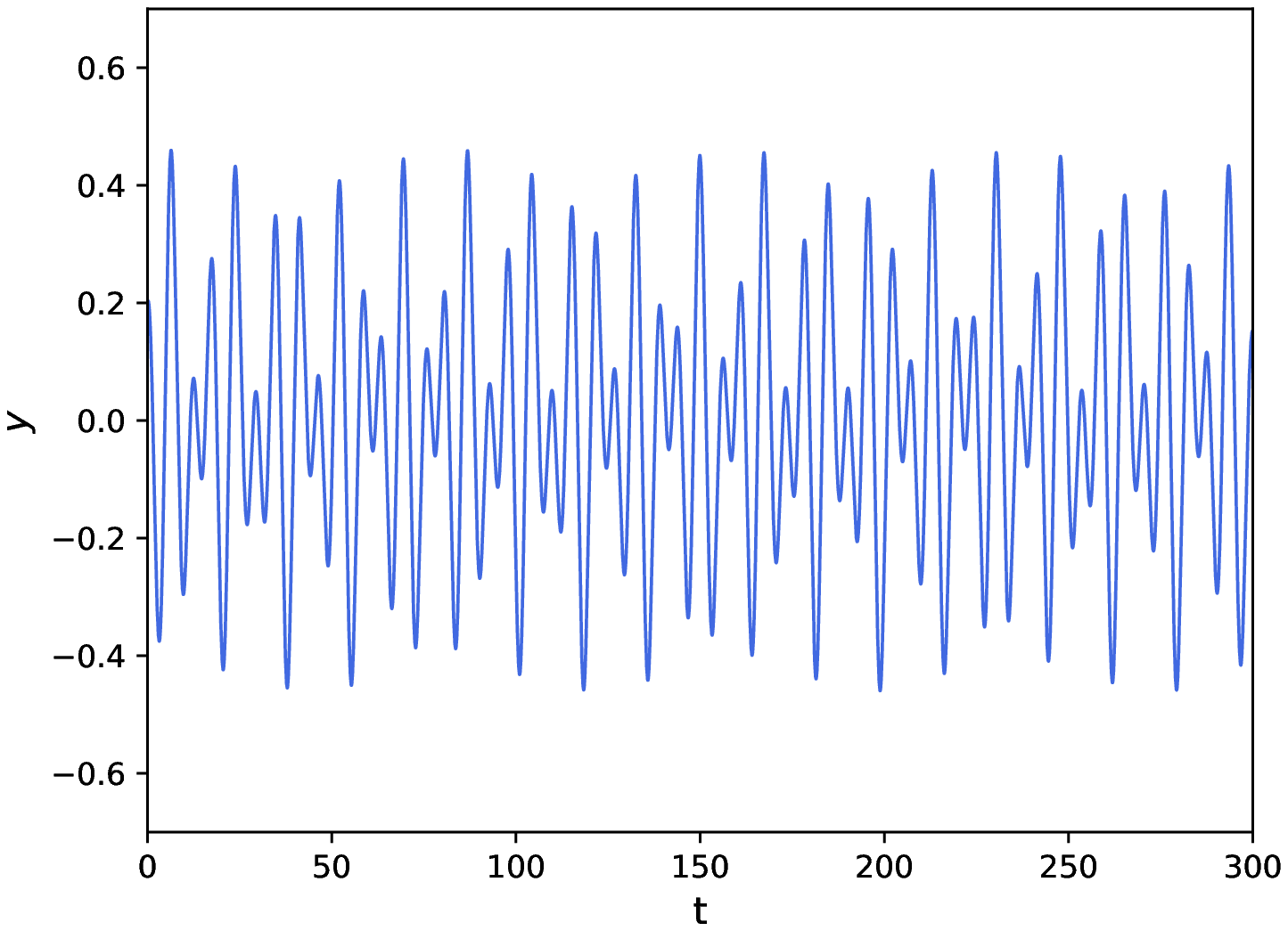}
\caption{$\alpha=-1, \beta=.5, \Gamma=.2$}
\label{ytimeseries2}
\end{subfigure}%
\caption{(Color online) \ Regular solutions of Eq. (\ref{duff-eqn}) in the vicinity of the point
$P_0$ with the initial conditions $x(0)=.1, y(0)=0.2, \dot{x}(0)=.03$ and $\dot{y}(0)=.04$.}
\label{time-series-P0}
\end{figure}

\begin{figure}[htbp]
\begin{subfigure}{.5\textwidth}
\centering
\includegraphics[width=.8\linewidth]{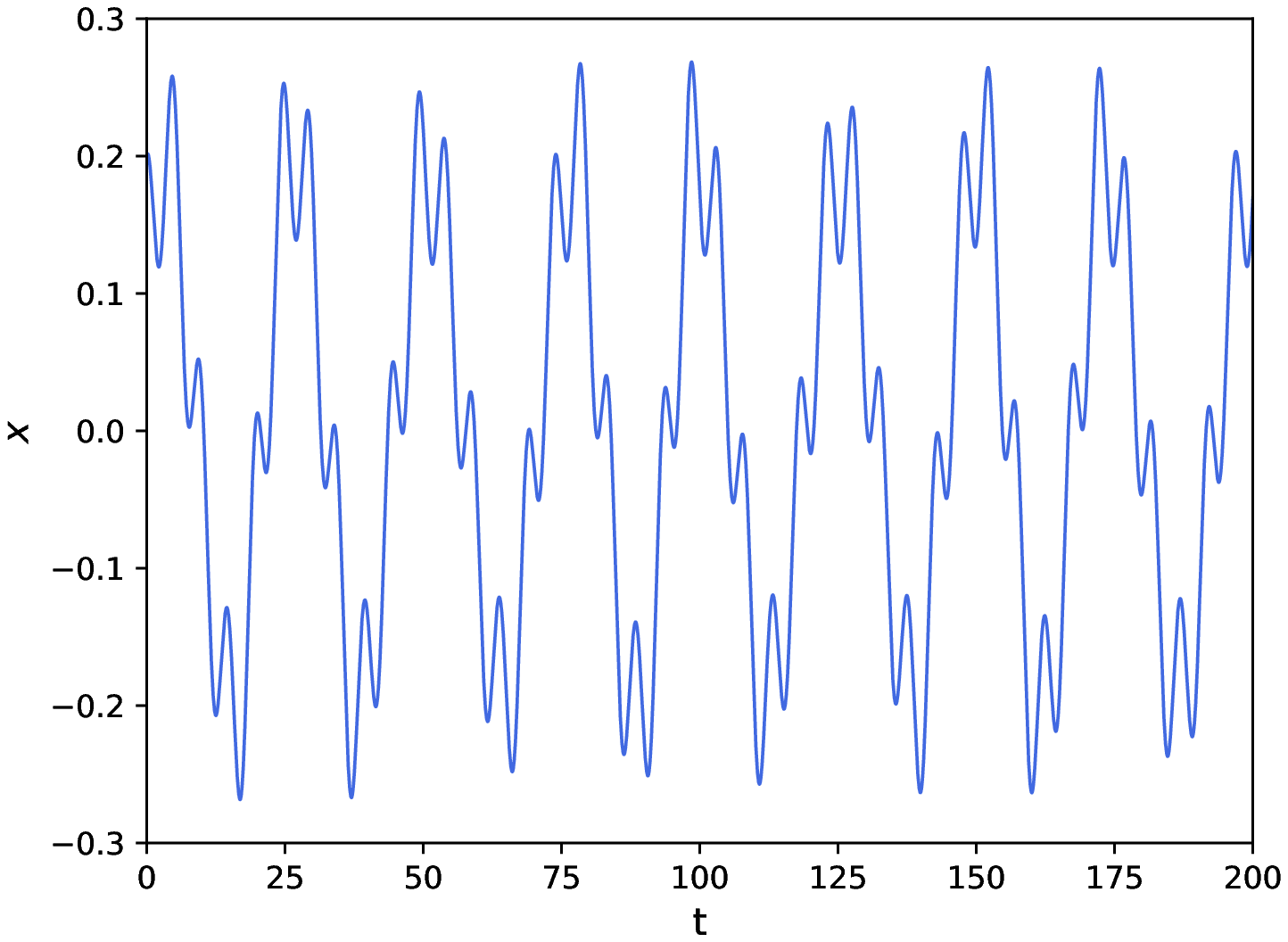} 
\caption{ $\alpha=1, \beta=1.01, \Gamma=.3$}
\label{xtimeseries3}
\end{subfigure}%
\begin{subfigure}{.5\textwidth}
\centering
\includegraphics[width=.8\linewidth]{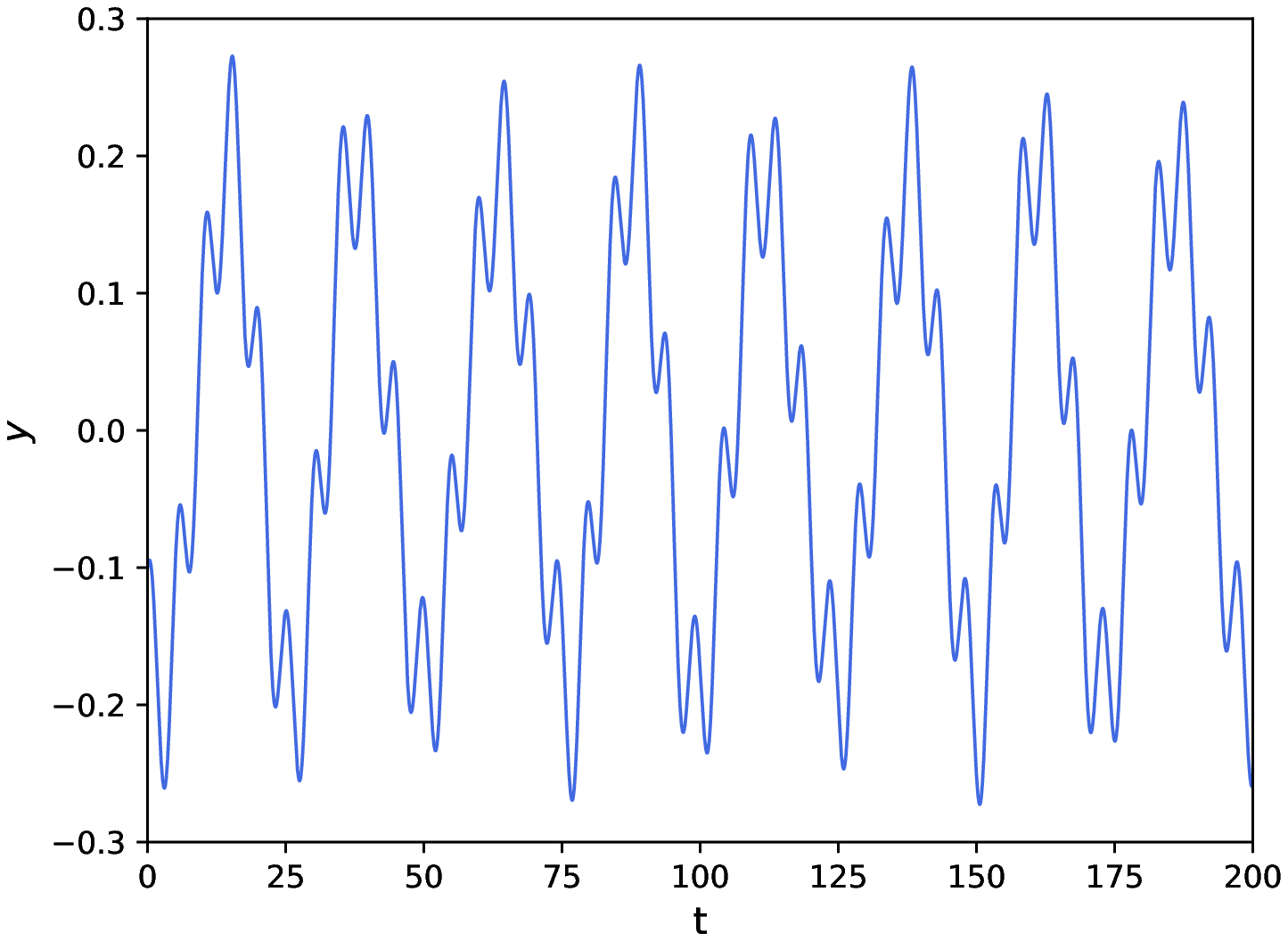}  
\caption{$\alpha=1, \beta=1.01, \Gamma=.3$}
\label{ytimeseries3}
\end{subfigure}%
\caption{(Color online) \ Regular solutions of Eq. (\ref{duff-eqn}) in the
vicinity of the point $P_1^{+}$ with the initial conditions $x(0)=.2, y(0)=-0.1, \dot{x}(0)=.02$
and $\dot{y}(0)=.03$.}
\label{time-series-P1}
\end{figure}

The time-series of the dynamical variables  in the vicinity of the point
$P_0$ is shown in Fig. (\ref{time-series-P0}) for $\Gamma=.2, \beta=.5$ and 
$\alpha=\pm 1$. Periodic solutions in Figs.(\ref{xtimeseries1}) and
(\ref{ytimeseries1}) correspond to $\alpha=1$, while Figs.
(\ref{xtimeseries2}) and (\ref{ytimeseries2}) correspond to $\alpha=-1$.
It may be noted that the time evolution of the dynamical variables
with the same initial conditions and fixed $\Gamma, \beta$ show similar
oscillatory behaviour for positive as well as negative $\alpha$.
There are minute changes in amplitudes and phases and that too in the limit of
large $t$. It has been checked numerically that the same feature also persists
for smaller as well as higher values of $\alpha$. The periodic solutions in
the vicinity of the points $P_1^{\pm}$ exist only for $\alpha >0$, confirming
the results of the linear stability analysis. The solutions around $P_1^{+}$
are shown in Fig. (\ref{time-series-P1}) for $\alpha=1, \beta=1.01, \Gamma=.3$.
The Lyapunov exponents and the autocorrelation
functions for the time series representing the periodic solutions in Fig. (\ref{time-series-P1})
have been calculated to confirm that these solutions are indeed regular. 
The transition from regular to chaotic behaviour is seen as $\beta$ is increased beyond $\tilde{\beta}_c
\sim 1.1$ under similar conditions, i.e.  $\alpha=1, \gamma=.3$ and  $x(0)=.2, y(0)=-0.1, \dot{x}(0)=.02$
and $\dot{y}(0)=.03$. It may be noted that the initial conditions for the bifurcation diagram in Fig.-2
is different from the initial conditions used for periodic solutions around the point $P_1$ in Fig.-4. Thus,
the critical value of $\beta_c$ is different for the two cases. The equilibrium points
\begin{figure}[ht!]
\begin{subfigure}{.5\textwidth}
\centering
\includegraphics[width=.8\linewidth]{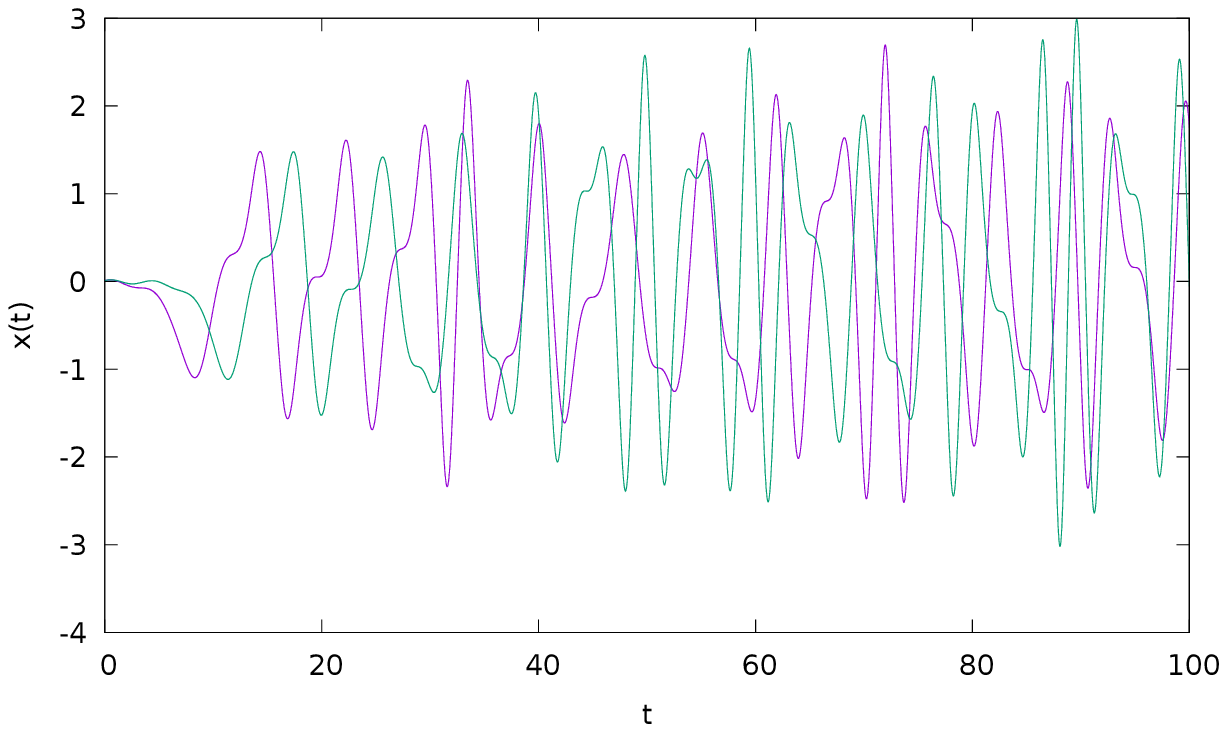} 
\caption{ $\Gamma=0.01, \beta=1.5, \alpha=.5$}
\label{chaotic1}
\end{subfigure}%
\begin{subfigure}{.5\textwidth}
\centering
\includegraphics[width=.8\linewidth]{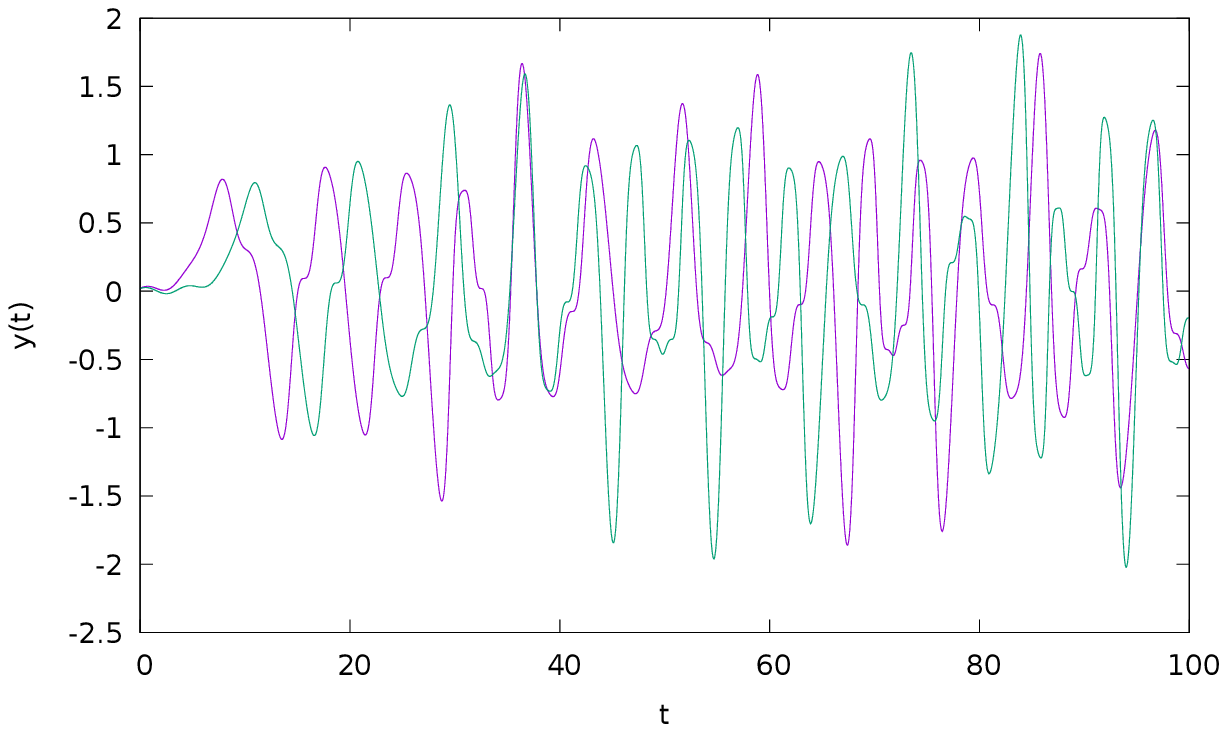}  
\caption{$\Gamma=0.01, \beta=1.5, \alpha=.5$}
\label{chaotic2}
\end{subfigure}%
\newline
\begin{subfigure}{.5\textwidth}
\centering
\includegraphics[width=.8\linewidth]{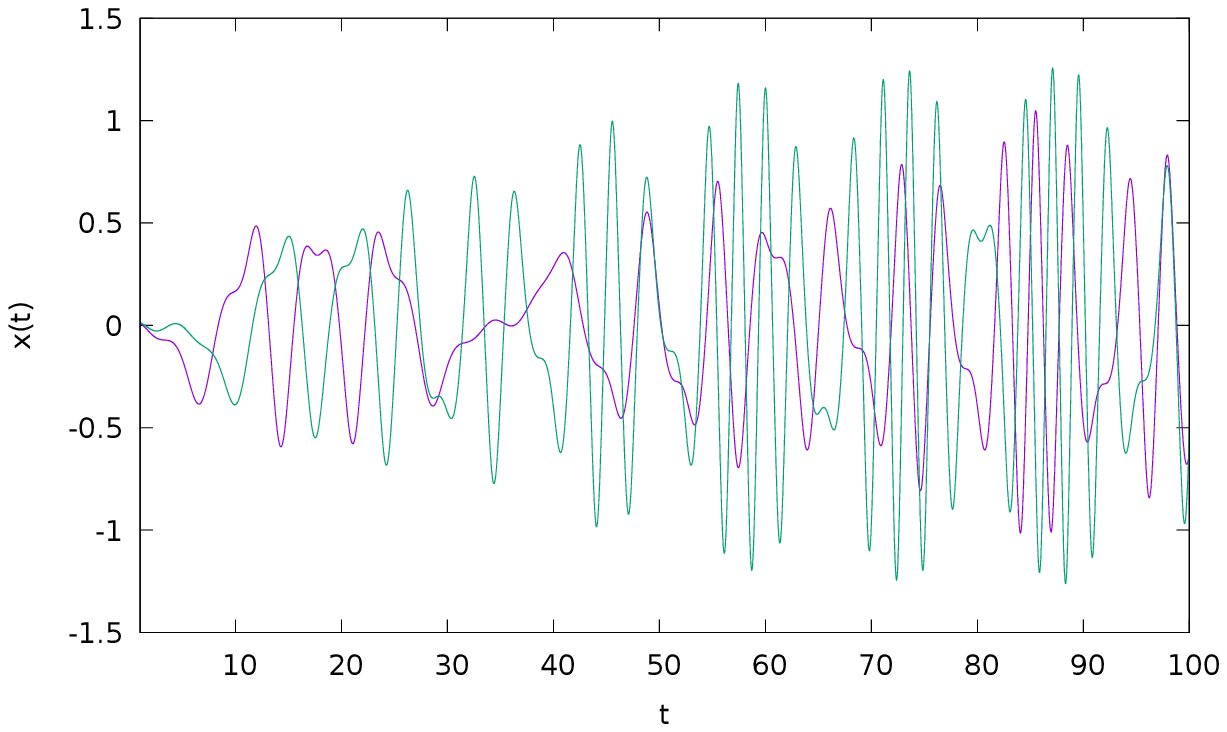}
\caption{$\Gamma=0.01, \beta=1.5, \alpha=5$}
\label{chaotic3}
\end{subfigure}%
\begin{subfigure}{.5\textwidth}
\centering
\includegraphics[width=.8\linewidth]{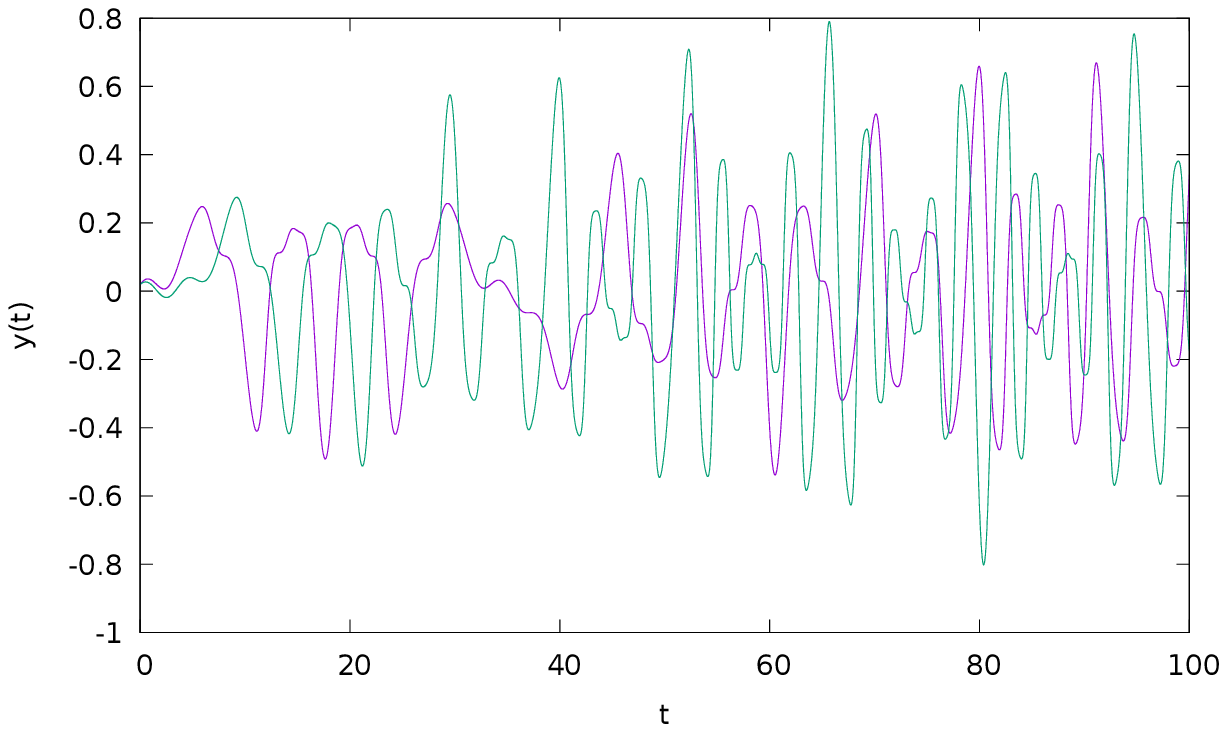}
\caption{$\Gamma=0.01, \beta=1.5, \alpha=5$}
\label{chaotic4}
\end{subfigure}%
\caption{(Color online) Chaotic solutions of Eq. (\ref{duff-eqn}) with two sets of initial
conditions (a) $x(0)=.01, y(0)=0.02, \dot{x}(0)=.03, \dot{y}(0)=.04$ (violet color) and 
(b) $x(0)=.01, y(0)=0.02, \dot{x}(0)=.03, \dot{y}(0)=.025$ (green colour)
} 
\label{chaotic}
\end{figure}
\noindent $P_1^{\pm}$ are related to each other as $P_1^-=-P_1^+$ for fixed $\alpha$ and
$\beta$. Further, Eq. (\ref{duff-eqn}) remains invariant under $x\rightarrow -x, y \rightarrow -y$. Thus, the
solutions around the equilibrium point $P_1^{-}$ for the same values of the parameters $\alpha=1, \beta=1.01,
\gamma=.3$, but with the initial conditions $x(0)=-.2, y(0)=0.1, \dot{x}(0)=-.02$ and $\dot{y}(0)=-.03$ may simply
be obtained by taking mirror image of the plot in Fig. 4.(a) with respect to $x=0$ and  in Fig. 4.(b) with respect
to $y=0$. No separate numerical solution around $P_1^-$ is presented for this reason.
\subsection{Chaotic Dynamics}
\begin{figure}[ht!]
\begin{subfigure}{.5\textwidth}
\centering
\includegraphics[width=.8\linewidth]{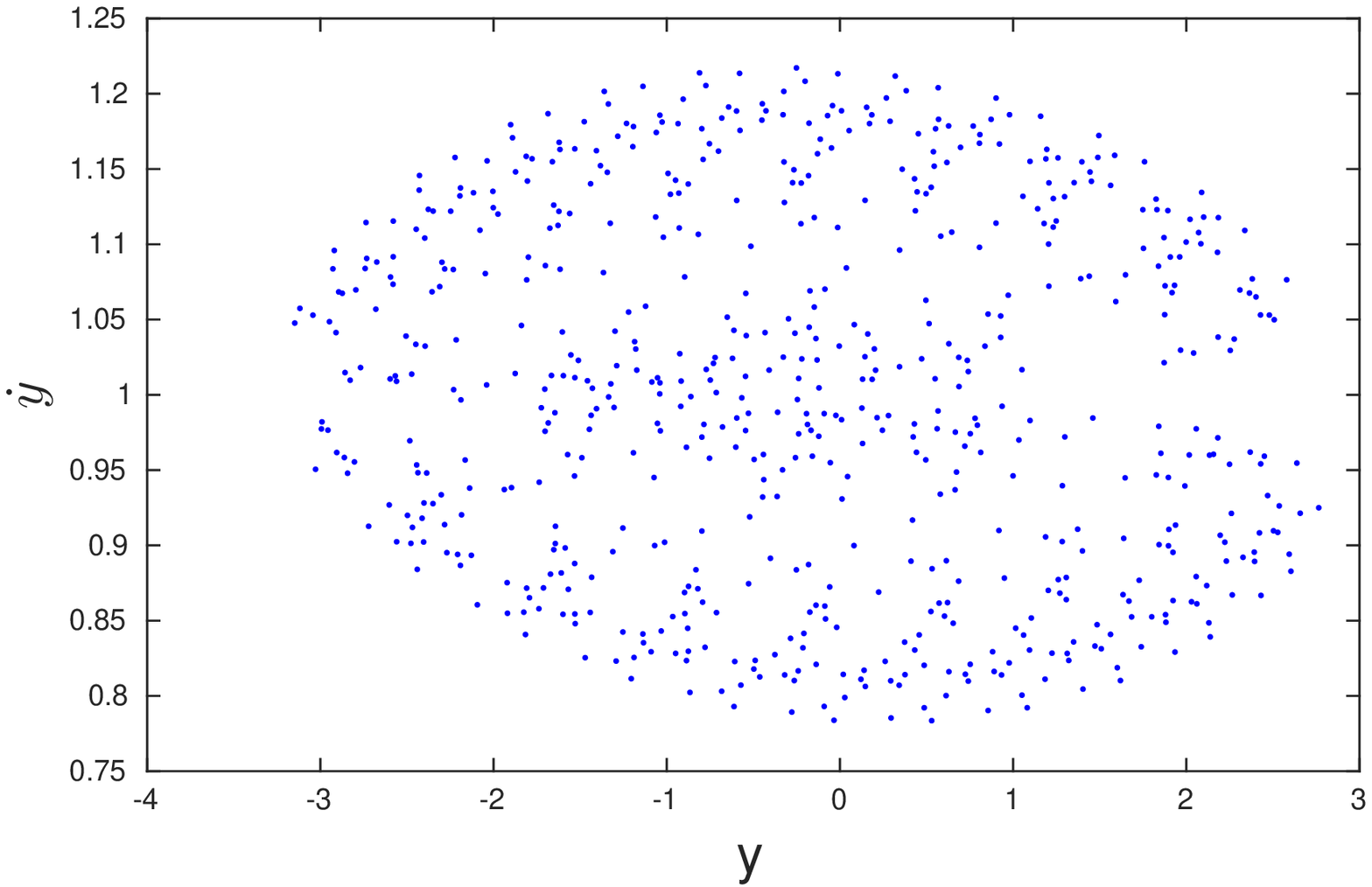}
\caption{Poincar$\acute{e}$ section: $\dot{y}(t)$ VS. $y(t)$ plot}
\label{multi-poincare}
\end{subfigure}%
\begin{subfigure}{.5\textwidth}
\centering
\includegraphics[width=.8\linewidth]{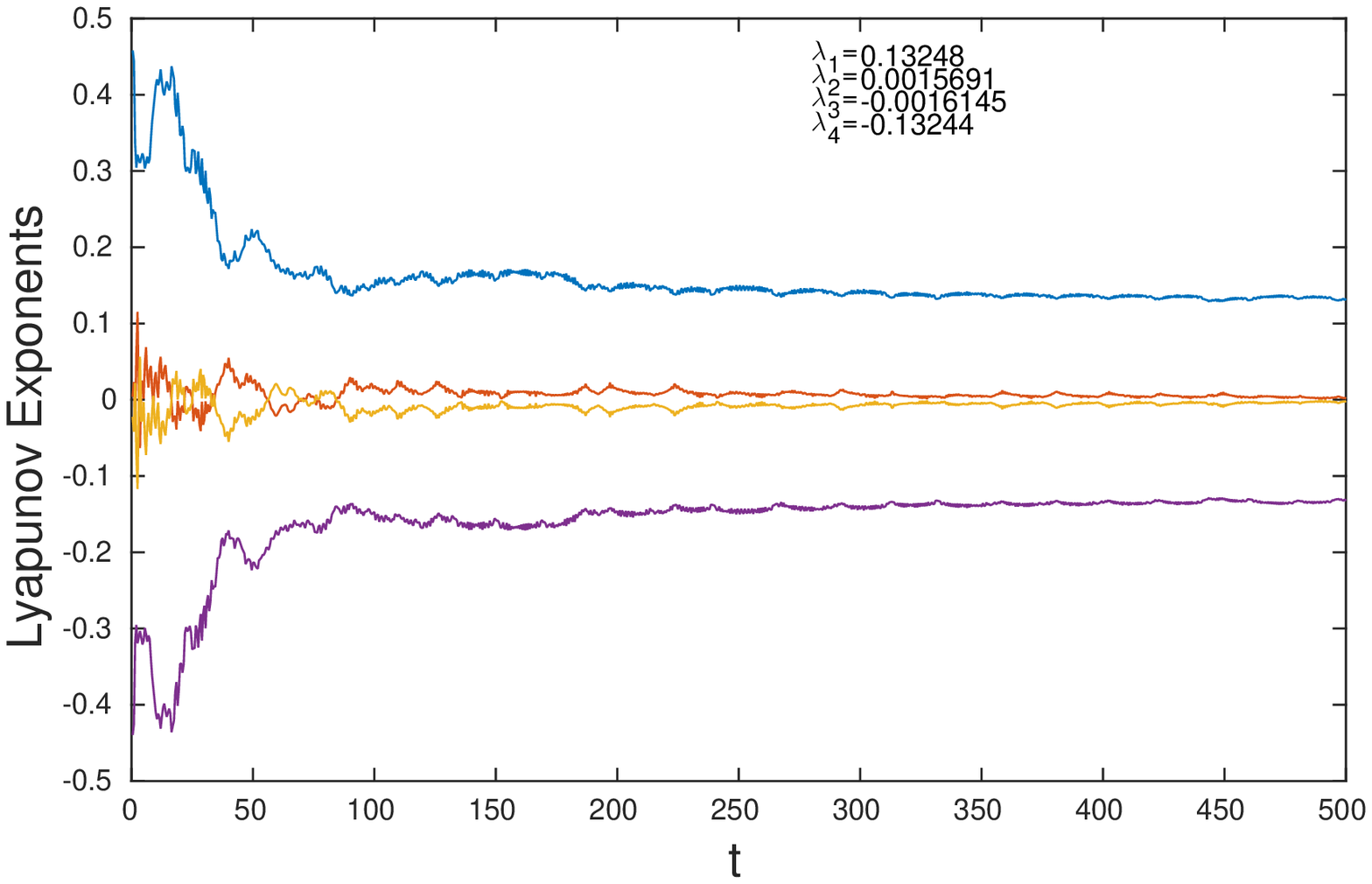}
\caption{Lyapunov exponents}
\label{multi-lyap}
\end{subfigure}%
\newline
\begin{subfigure}{.5\textwidth}
\centering
\includegraphics[width=.8\linewidth]{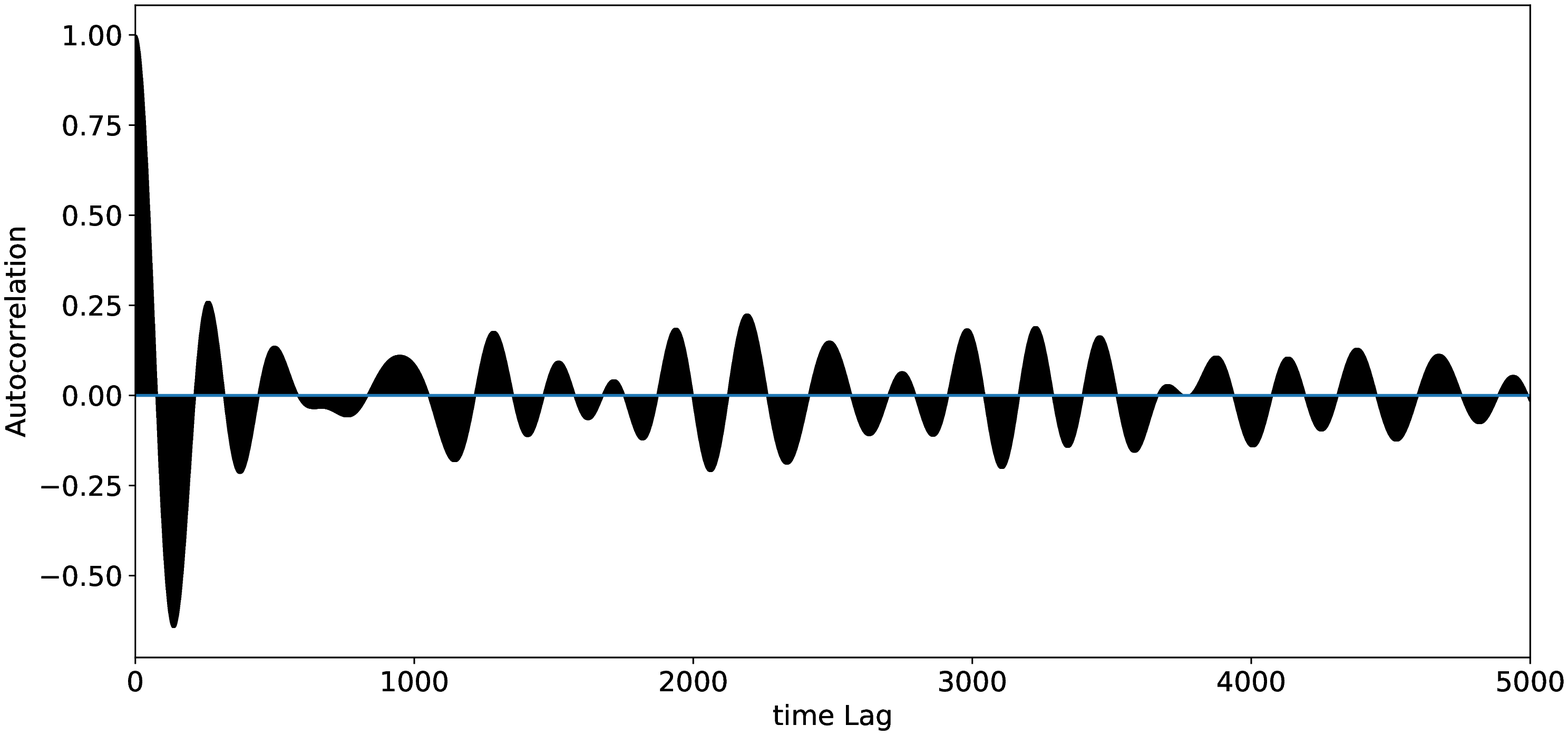}
\caption{Autocorrelation function of $x(t)$}
\label{multi-xauto}
\end{subfigure}%
\begin{subfigure}{.5\textwidth}
\centering
\includegraphics[width=.8\linewidth]{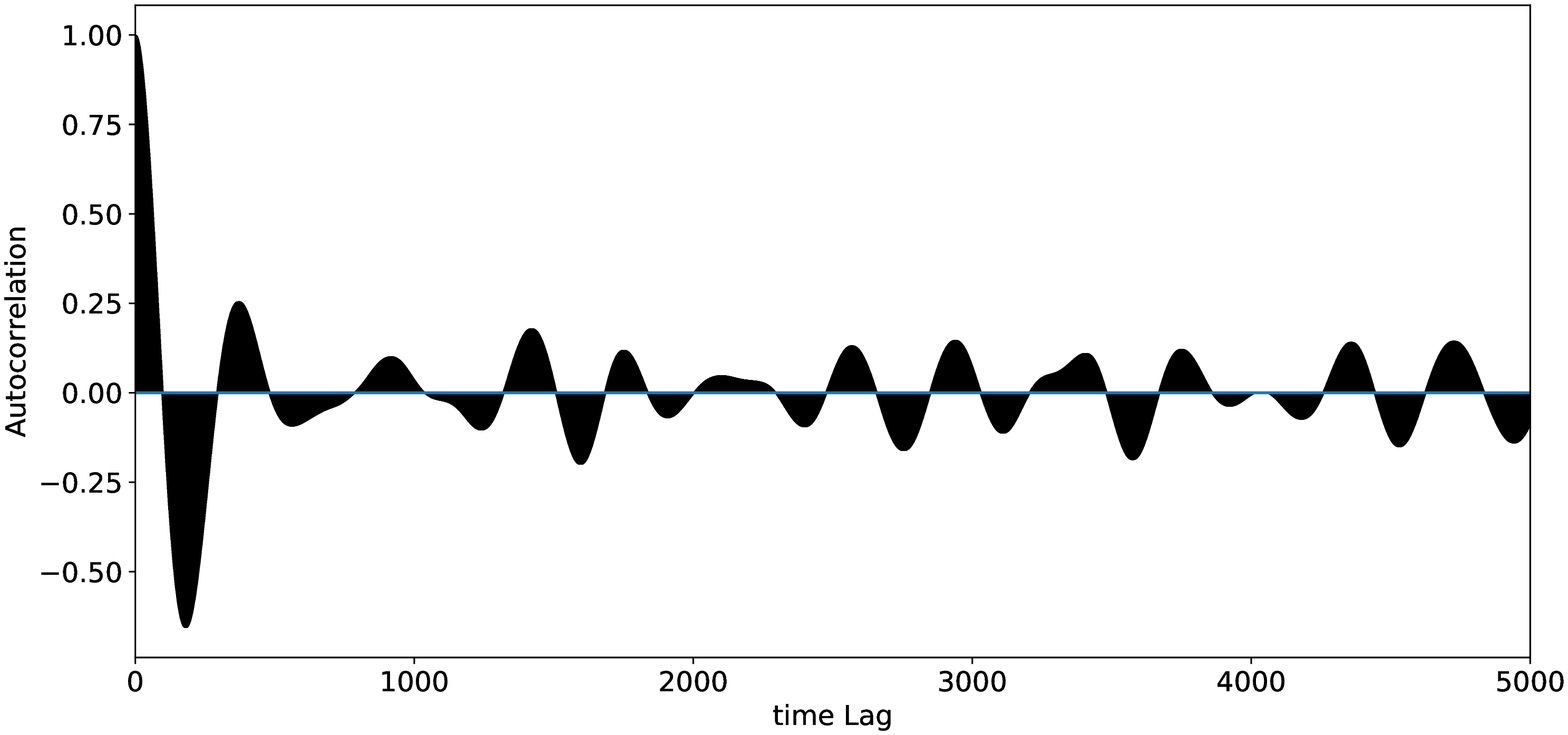}
\caption{Autocorrelation function of $y(t)$}
\label{multi-yaoto}
\end{subfigure}%
\newline
\begin{subfigure}{.5\textwidth}
\centering
\includegraphics[width=.8\linewidth]{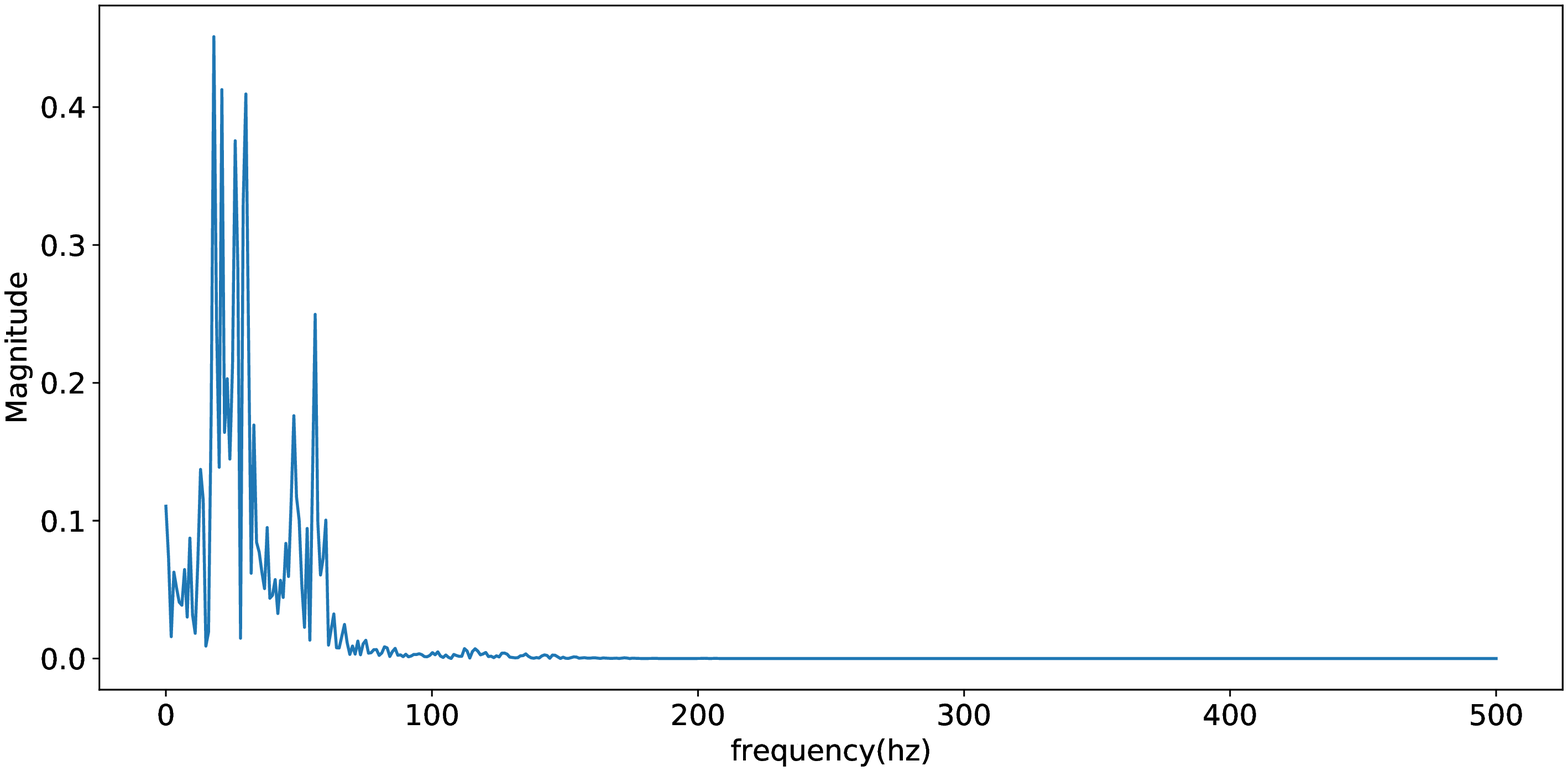}
\caption{Powerspectra of $x(t)$}
\label{multi-xpower}
\end{subfigure}%
\begin{subfigure}{.5\textwidth}
\centering
\includegraphics[width=.8\linewidth]{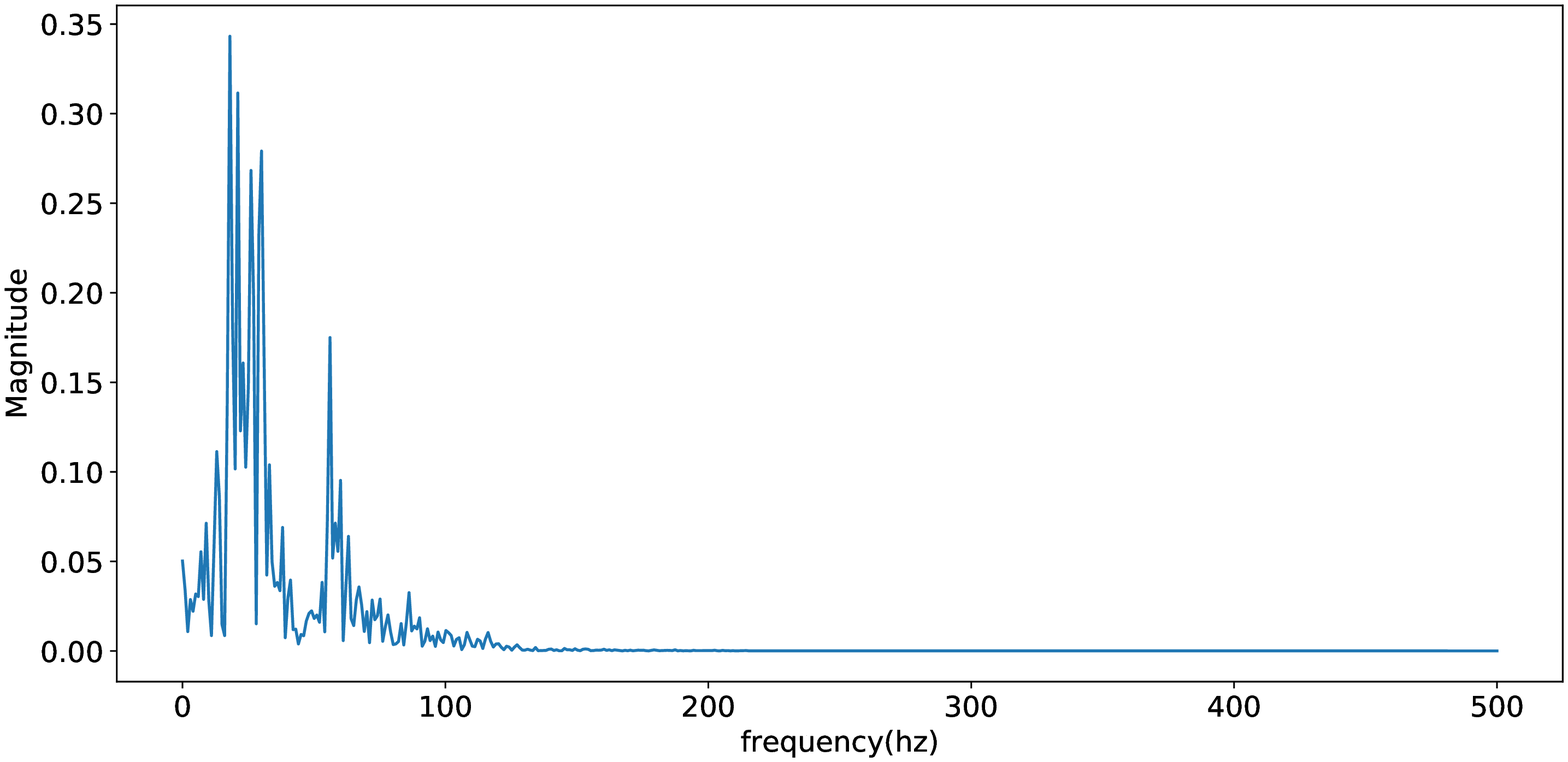}
\caption{Powerspectra of $y(t)$}
\label{multi-ypower}
\end{subfigure}%
\caption{(Color online)  Poincar$\acute{e}$ section, Lyapunov exponents, autocorrelation function
and power spectra for $\Gamma=0.01, \beta=1.5, \alpha=.5$ with the initial condition
$x(0)=.01, y(0)=0.02, \dot{x}(0)=.03, \dot{y}(0)=.04$ }
\label{multi}
\end{figure}

The bifurcation diagram shows that the system with $\Gamma=0.01$ and $\alpha=.5$ is chaotic for $\beta > \beta_c=1.05$. The sensitivity
of the dynamical variables to the initial conditions are studied in different regions of the parameters by considering two sets
of initial conditions: (a) $x(0)=.01, y(0)=0.02, \dot{x}(0)=.03, \dot{y}(0)=.04$ and
(b) $x(0)=.01, y(0)=0.02, \dot{x}(0)=.03, \dot{y}(0)=.025$.  These two initial conditions are identical except for the values
of $\dot{y}(0)$ which differ by $.015$. The time series of the dynamical variables in the chaotic regime is presented in Fig. \ref{chaotic} for $\beta=1.5$.
The chaotic behaviour in the model has been confirmed by other independent methods also. In this regard,
the auto-correlation function, Lyapunov exponent, Poincar$\acute{e}$ section and power spectra are plotted in Fig.  (\ref{multi}).
\noindent The Lyapunov exponents are $(.13248, .0015691, -.0016145, -.13244)$. The sum of the Lyapunov
exponents are zero for a Hamiltonian system which is valid for the present case if values up to the third decimal
places are considered with an error of the order of $10^{-4}$. 

\begin{figure}[ht!]
\begin{subfigure}{.5\textwidth}
\centering
\includegraphics[width=.8\linewidth]{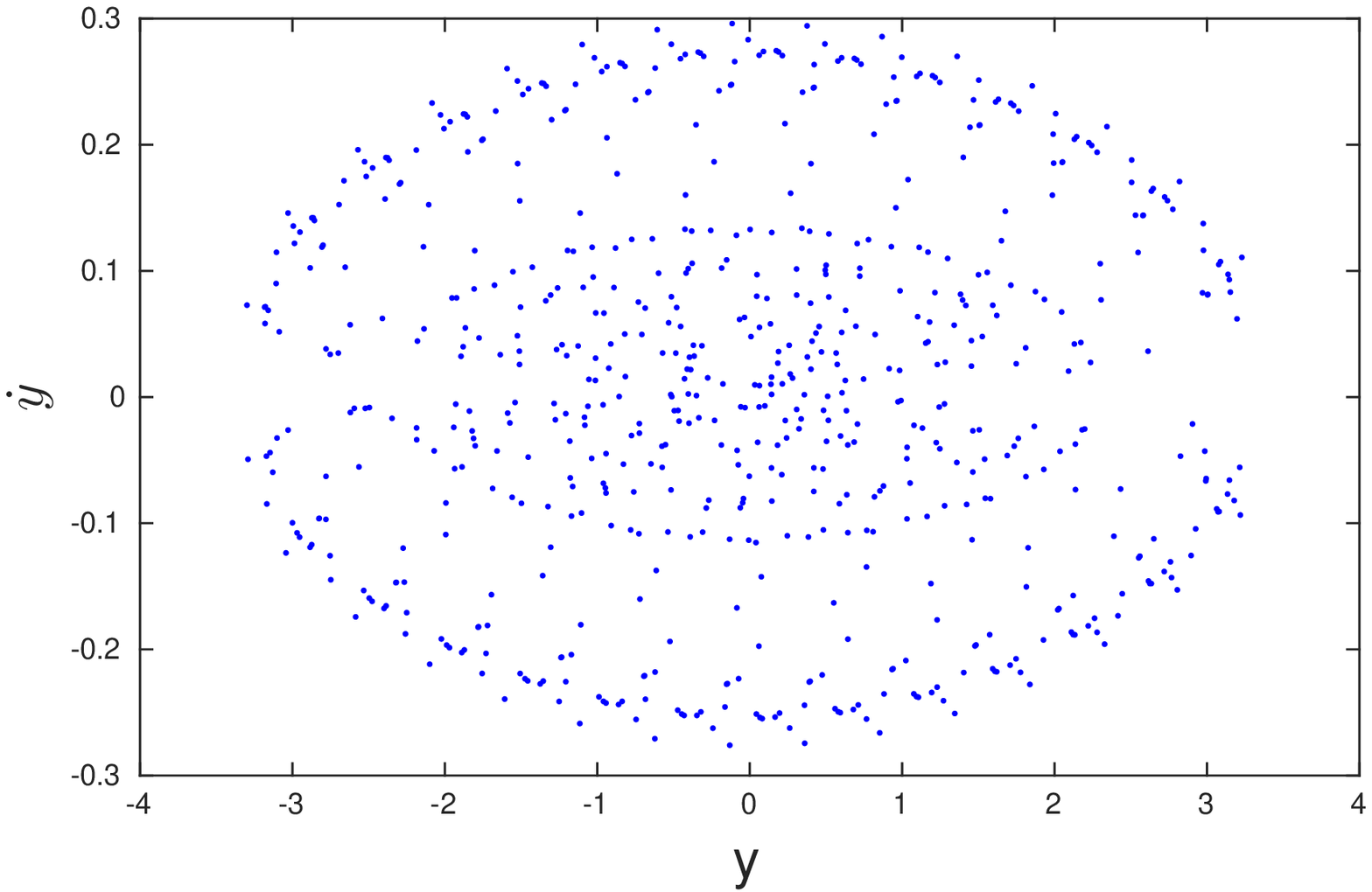}
\caption{Poincar$\acute{e}$ section: $\dot{y}(t)$ VS. $y(t)$ plot}
\label{ps-gamma0}
\end{subfigure}%
\begin{subfigure}{.5\textwidth}
\centering
\includegraphics[width=.8\linewidth]{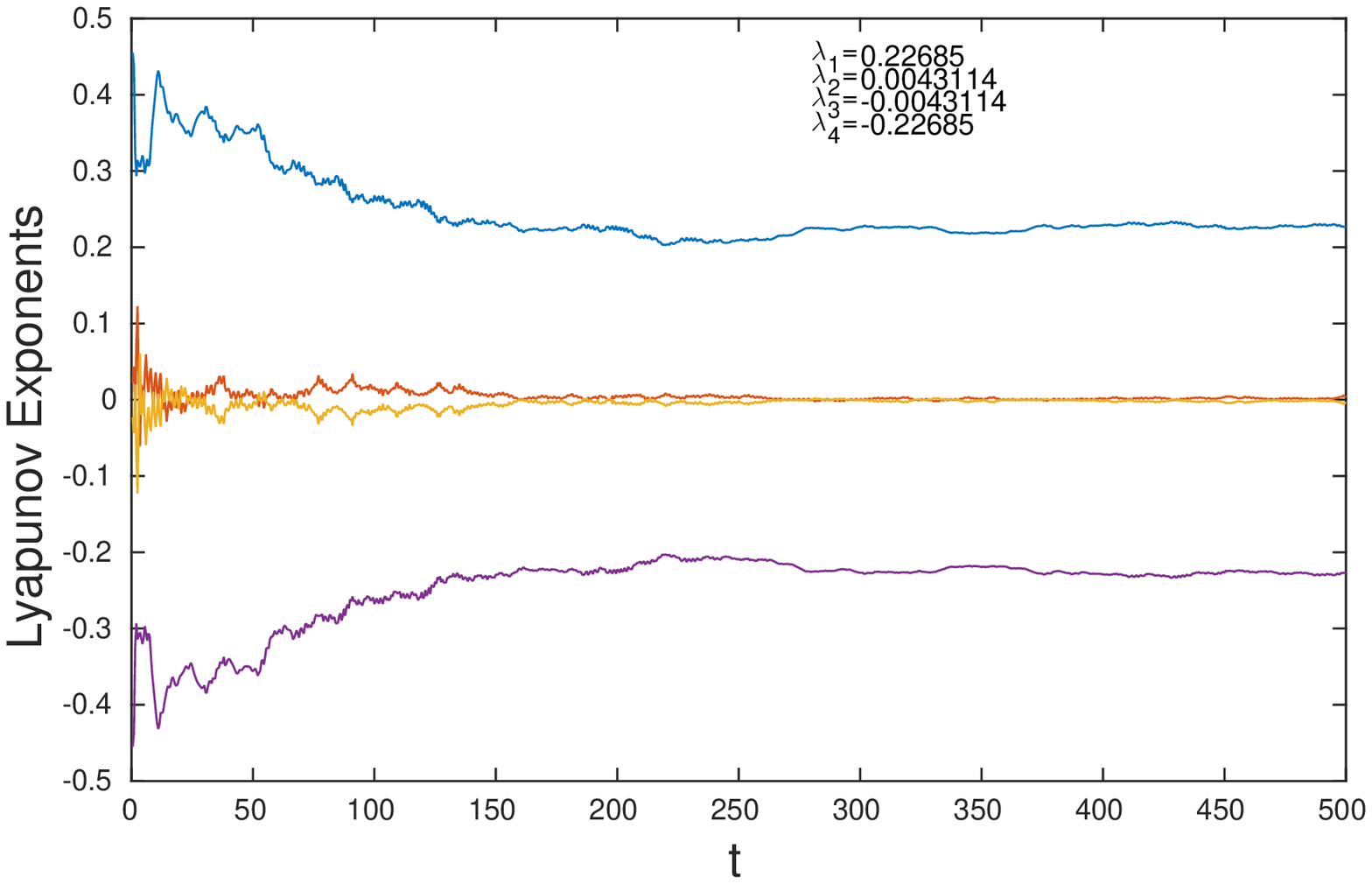}
\caption{Lyapunov exponents}
\label{g0-lyapunov}
\end{subfigure}%
\newline
\begin{subfigure}{.5\textwidth}
\centering
\includegraphics[width=.8\linewidth]{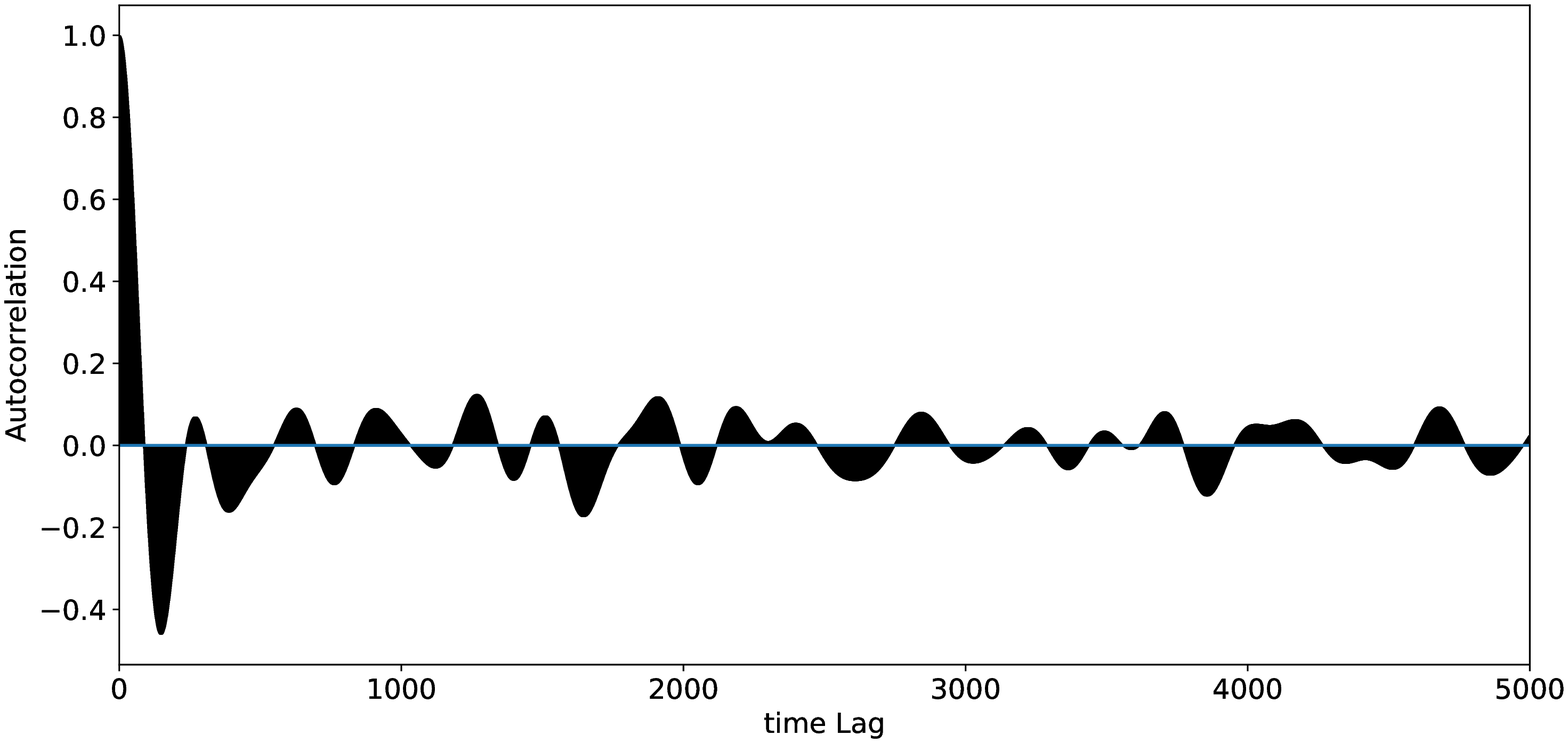}
\caption{Autocorrelation function of $x(t)$}
\label{acf-x-g0}
\end{subfigure}%
\begin{subfigure}{.5\textwidth}
\centering
\includegraphics[width=.8\linewidth]{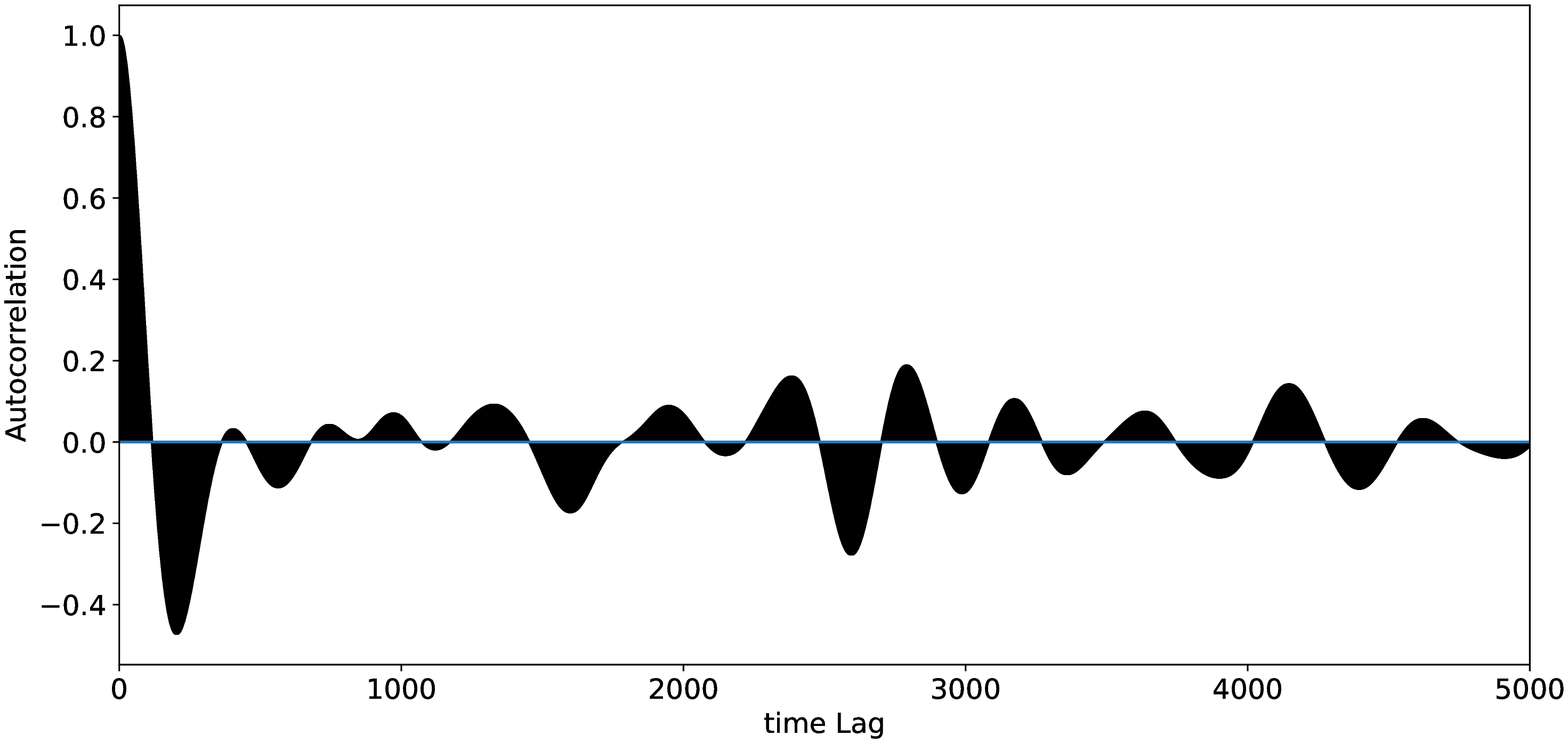}
\caption{Autocorrelation function of $y(t)$}
\label{acf-y-g0}
\end{subfigure}%
\newline
\begin{subfigure}{.5\textwidth}
\centering
\includegraphics[width=.8\linewidth]{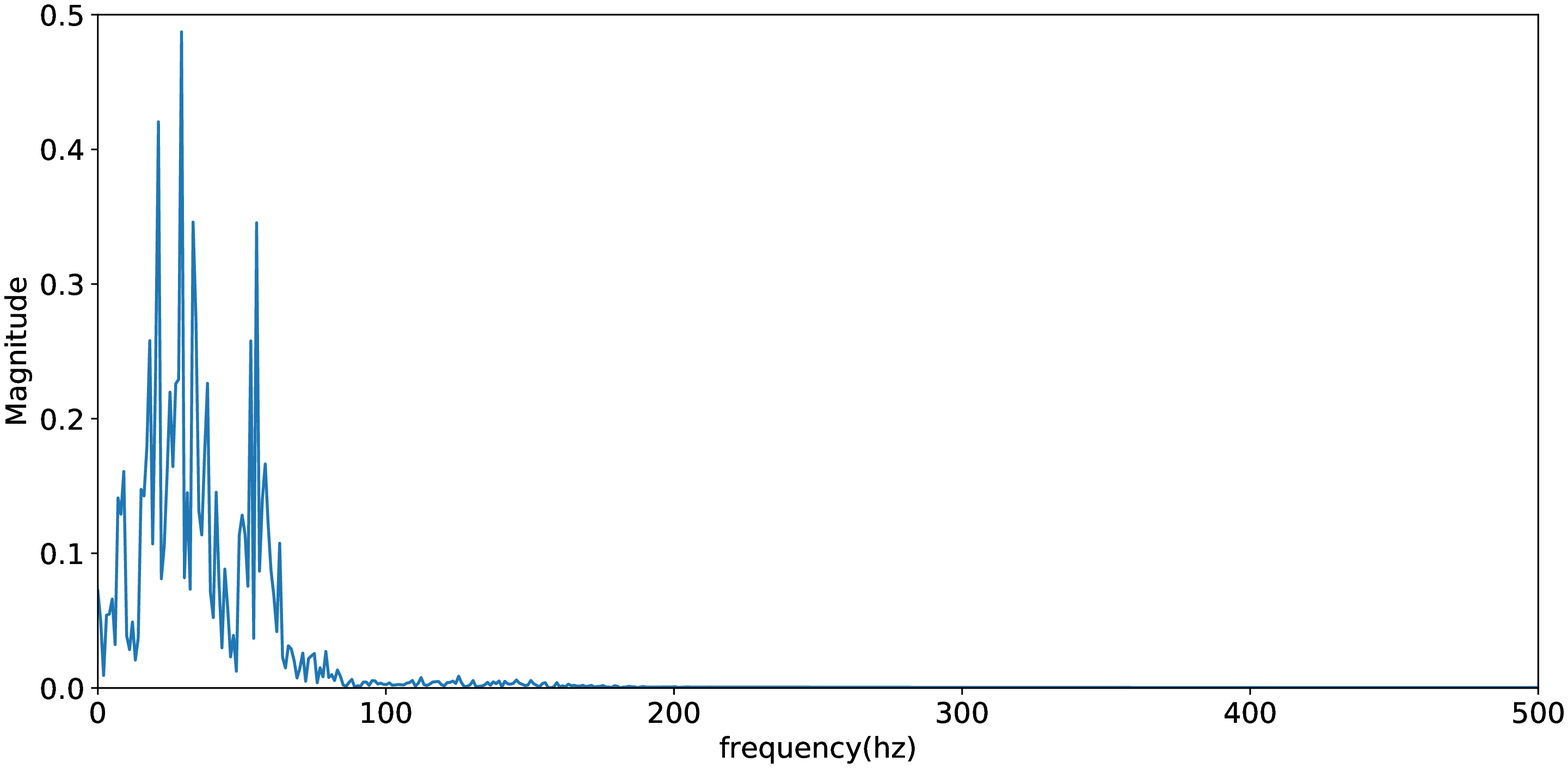}
\caption{Powerspectra of $x(t)$}
\label{g0-psx}
\end{subfigure}%
\begin{subfigure}{.5\textwidth}
\centering
\includegraphics[width=.8\linewidth]{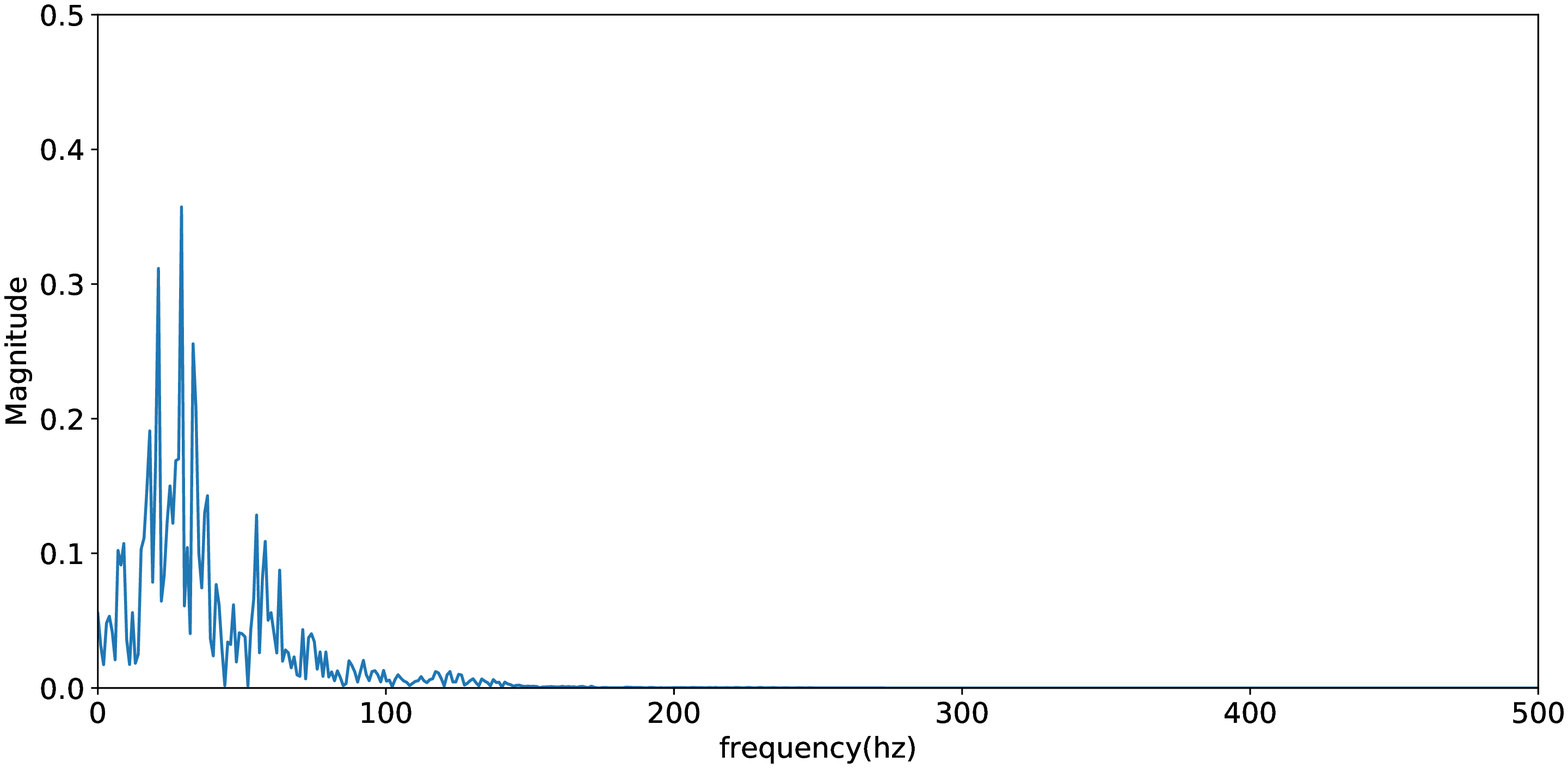}
\caption{Powerspectra of $y(t)$}
\label{g0-psy}
\end{subfigure}%
\caption{(Color online) Poincar$\acute{e}$ section, Lyapunov exponents, autocorrelation functions and
power spectra for $\Gamma=0, \alpha=.5, \beta=1.5$ and the initial conditions $x(0)=.01, y(0)=.02,
\dot{x}(0)=.03, \dot{y}(0)=.04$} 
\label{chaos-gamma0}
\end{figure}

The main emphasis of this article is on Hamiltonian system with balanced loss and gain. However, the
Hamiltonian with $\Gamma=0$, i.e. no gain-loss regime,  deserves special attention due to its rich
dynamical properties. For $\Gamma=0$, the point $P_0$ is stable for $0 < \beta^2 < 1$, while the points
$P_1^{\pm}$ are stable for $\beta^2 >1$. Further, the quantities $Q_i$'s are purely imaginary and
Eq. (\ref{t1-eqn}) is exactly solvable without any assumptions on $Q_i$'s. It has been checked numerically
that periodic solutions of Eq. (\ref{duff-eqn}) exist for $\Gamma=0$ near the equilibrium points
$P_0$, $P_1^{\pm}$. The most important result is that the Hamiltonian $\tilde{H}$ with $\Gamma=0$
is chaotic. The Poincar$\acute{e}$ section, Lyapunov exponents, autocorrelation functions and power spectra
for $\Gamma=0, \alpha=.5, \beta=1.5$ and the initial conditions
$x(0)=.01, y(0)=.02, \dot{x}(0)=.03, \dot{y}(0)=.04$ are presented in Fig.-\ref{chaos-gamma0}. The
Lyapunov exponents are $(0.22685, 0.0043114, -.0043114, - 0.22685)$ which may be used to compute other
measures of a physical system. The highest Lyapunov exponent for $\Gamma=0$ is greater than the highest
Lyapunov exponent for $\Gamma=.01$ with all other conditions remaining the same.
\noindent It may be noted that the Duffing oscillator admits chaotic solutions provided both damping
and external driving terms are present. However, for the case of coupled Duffing oscillator model of
this article, chaotic behaviour is observed for the system without any loss-gain terms. The coupling
to the linear oscillator with $x$-dependent angular frequency provides driving force to the Duffing oscillator.
The Hamiltonian $H_u$ in Eq. (\ref{u-hami}) takes a simple form for $\gamma=0, \beta_1=\beta_2$:
\bea
H_u  =  \left ( \frac{1}{2} P_u^2+ \frac{\Omega_+}{2} u^2 +\frac{g}{4} u^4 \right ) -
\left ( \frac{1}{2} P_v^2 + \frac{\Omega_-}{2} v^2 + \frac{g}{4} v^4 \right )
+ \frac{g \ u v}{2} \left ( u^2-v^2 \right ),  \Omega_{\pm} =\omega^2 \pm \beta_1.
\eea
\noindent describing two nonlinearly coupled undamped unforced Duffing oscillators.
This also provides an example of Hamiltonian chaos within a simple framework which deserves further
investigations. 

\section{Non-${\cal{PT}}$-symmetric Dimer with balanced loss and gain}

Different types of dimer models play an important role in many areas of physics
and in particular, in the context of ${\cal{PT}}$ symmetric
systems\cite{ivb,khare, kono,aac,wunner}. In this section, an exactly solvable
non-${\cal{PT}}$-symmetric Hamiltonian describing a dimer model with balanced
loss and gain is shown to admit periodic solutions. A standard route to
the occurrence of dimer models is via different approximation methods,
including the multiple time scale analysis.
The time-evolution of the amplitude in the leading order of the perturbation
is described by dimer models. For the case of the coupled Duffing oscillator
model, the resulting dimer models for $\alpha \ll 1$ in Eq. (\ref{t1-eqn})
and for $\gamma \ll 1, \alpha \ll 1$ in Eq. (\ref{t3-eqn}) do not contain any
loss-gain terms. A dimer model with balanced loss and gain is obtained 
for $\Gamma \ll 1, \beta \ll 1$ that is described in this section.

Both $\Gamma$ and $\beta$ are treated as small parameters with the
identification of $\Gamma \equiv \epsilon^2 \Gamma_0,
\beta \equiv \epsilon^2 \beta_0, \ \epsilon  \ll 1$. The strength of the
nonlinear interaction $\alpha$ is kept arbitrary. The time scales and
power series expansion of the space co-ordinates in terms of $\epsilon$
are chosen as,
\bea
T_0=t, T_{2n} = \epsilon^{2n} t, n=1, 2, \dots, \ \
X=\sum_{n=0}^{\infty} \epsilon^{2n+1} X^{(2 n+1)},
\label{var-scale-2}
\eea
\noindent so that there is no contribution from the nonlinear interaction
in the lowest order. In particular, after using Eq. (\ref{var-scale-2}), the
lowest two orders of the equations of motion (\ref{col-eqn}) read,
\bea
\label{epo1}
&& {\cal{O}}(\epsilon): \ \ \frac{\partial^2 X^{(1)}}{\partial T_0^2} +
X^{(1)} =0,\\
&& {\cal{O}}(\epsilon^3): \ \ \frac{\partial^2 X^{(3)}}{\partial T_0^2} +
X^{(3)} + 2  \frac{\partial^2 X^{(1)}}{\partial T_0 \partial_2} + 2 \Gamma_0
\sigma_3 \frac{\partial {X}^{(1)}}{\partial T_0} +
\beta_0 \sigma_1 X^{(1)} + \alpha
\bp
{x_{1}}^3\\ 3 {x_{1}}^2 y_{1} \ep =0.
\label{epo3}
\eea
\noindent The ${\cal{O}}(\epsilon)$ equation describes two decoupled isotropic
harmonic oscillators and the general solution has the form 
\bea
X^{(1)} = A(T_2, T_4,\dots) e^{i T_0} + c.c., \ \
A \equiv \begin{pmatrix} A_1(T_2,\dots) \\ B_1(T_2,\dots) \end{pmatrix},
\label{x1ep}
\eea
\noindent where the amplitude $A$ depends on slower time scales. The
time-evolution of $A$ as a function of $T_2$ is determined by the
equation\footnote{ A detailed derivation is skipped in order to avoid
repetition. The necessary steps as outlined in Sec. 3 for the case of
$\alpha \ll 1 $ may be followed to derive this equation.},
\bea
2 i \frac{\partial A}{\partial T_2}
+ \left ( 2 i \Gamma_0 \sigma_3 + \beta_0 \sigma_1 \right ) A
+ 3 \alpha \begin{pmatrix} {\vert A_1 \vert}^2 A_1\\
2 {\vert A_1 \vert}^2 B_1 + A_1^2 B_1^* \end{pmatrix}=0,
\label{dimer-blg}
\eea
\noindent which is obtained by substituting $X^{(1)}$ in Eq. (\ref{epo3})
and removing the secular terms. Eq. (\ref{dimer-blg}) describes a coupled
dimer model with nonlinear interaction. The amplitudes $A_i$'s may be
identified as wave propagating along the $i^{th}$ wave-guide.

Eq. (\ref{dimer-blg}) may also be derived from
the Lagrangian ${\cal{L}}$ or the corresponding Hamiltonian ${\cal{H}}$,
\bea
&& {\cal{L}} = \frac{i}{2} \left (A^{\dagger} \sigma_1 \dot{A} -
\dot{A}^{\dagger} \sigma_1 A \right ) -2 \Gamma_0 A^{\dagger} \sigma_2 A -
\beta_0 A^{\dagger} A - 3 \alpha {\vert A_1 \vert}^2 \left ( A_1 B_1^* +
A_1^* B_1 \right ),\nonumber \\
&& {\cal{H}}= 2 \Gamma_0 A^{\dagger} \sigma_2 A + \beta_0 A^{\dagger} A +
3 \alpha {\vert A_1 \vert}^2 \left ( A_1 B_1^* + A_1^* B_1 \right ).
\eea
\noindent The canonically conjugate variables are $(A_1, i B_1^{*})$.
The Hamiltonian system describes a nonlinear Schr$\ddot{o}$dinger dimer with
balanced loss and gain. The parameter $\Gamma_0$ measures the strength of the
loss-gain, while $\alpha$ is the coupling of the nonlinear interaction. The
system differs with most of the previous studies on Hamiltonian dimer model
with balanced loss and gain in one respect, it is not ${\cal{PT}}$ symmetric.
The parity and time-reversal transformation are defined as,
${\cal{P}}: A \rightarrow \sigma_1 A, \ {\cal{T}}: T_2 \rightarrow - T_2, i
\rightarrow -i$. The Hamiltonian is ${\cal{PT}}$ symmetric for $\alpha=0$.
The nonlinear interaction breaks ${\cal{PT}}$ symmetry and ${\cal{H}}$ is
non-${\cal{PT}}$-symmetric for $\alpha \neq 0$. 

The Hamiltonian dimer is an integrable system. Introducing the Stokes
variables,
\bea
Z_a=2 A^{\dagger} \sigma_a A, \ R=2 A^{\dagger} A=
\sqrt{Z_1^2+Z_2^2+ z_3^2}, \ a=1, 2, 3,
\eea
\noindent it immediately follows from Eq. (\ref{dimer-blg}) that $Z_1$ is
the second integral of motion, i.e. $\dot{Z_1}=0$. The constant
$Z_1$ is fixed to its initial value at $T_2=0$, i.e.  $Z_1(0) \equiv C_1$.
The dimer equations are transformed into a set of coupled linear differential
equation,
\bea
\dot{Z} = N Z, \ \
Z \equiv
\begin{pmatrix} Z_2\\ Z_3\\ R 
\end{pmatrix}, \ \
N \equiv
\begin{pmatrix}
0 && a && b\\
-a && 0 && - c\\
b && - c && 0
\end{pmatrix},
\eea
\noindent where the parameters $a, b, c$ are defined as,
\bea
a=\beta_0 +\frac{3 \alpha C_1}{4}, \ b=\frac{3 \alpha C_1}{4}, \
c= 2 \Gamma_0.
\eea
\noindent One of the eigenvalues of the matrix $N$ is zero and the remaining
two eigenvalues are purely imaginary numbers and complex conjugate of each
other for real $\eta$:
\bea
\eta_1=0, \eta_2=i \eta, \eta_3=-i \eta, \ \
\eta \equiv \sqrt{a^2-b^2-c^2}=\sqrt{ \left ( \beta_0^2 + \frac{3}{2} \alpha 
\beta_0 C_1 \right ) - 4 \Gamma_0^2 }.
\eea
\noindent There are growing as well as decaying modes whenever $\eta$ becomes
imaginary. The parameter $\eta$ is real for the following condition,
\bea
\beta_0^2 + \frac{3}{2} \alpha \beta_0 C_1 - 4 \Gamma_0^2 \geq 0.
\label{real-con}
\eea
\noindent It is interesting to note that for a fixed set of parameters
the integration constant $C_1$ may be always chosen such that $\eta$
is real. The general solutions has the expression,
\bea
Z= C_2 \begin{pmatrix} -\frac{c}{a}\\ -\frac{b}{a}\\ 1 \end{pmatrix}
+ C_3 \begin{pmatrix} i \frac{-a(ab-i c\eta)+b(b^2+c^2)}{(b^2+c^2)\eta}\\
- \frac{ab-i c\eta}{b^2+c^2}\\1 \end{pmatrix} e^{-i \eta \epsilon^2 t}
+ C_4 \begin{pmatrix} i \frac{a(ab+i c\eta)-b(b^2+c^2)}{(b^2+c^2)\eta}\\
- \frac{ab+i c\eta}{b^2+c^2}\\1 \end{pmatrix} e^{i \eta \epsilon^2 t},
\eea
\noindent where $C_2, C_3, C_4$ are integration constants. The solutions
for $A_1$ and $A_2$ may be obtained as,
\bea
A_1=\frac{1}{2} \sqrt{R+Z_3} \ e^{i \theta_1}, \
A_2=\frac{1}{2} \sqrt{R-Z_3} \ e^{i \theta_2}, 
\label{inv-stokes-1}
\eea
\noindent where the phases are determined from the equations,
\bea
\frac{d \theta_1}{d T_2} = \frac{\beta_0}{2} \frac{Z_1}{R+Z_3}
+ \frac{3}{4} (R+Z_3), \ \
\theta_2=\theta_1+\tan^{-1}\frac{Z_2}{Z_1}.
\label{inv-stokes-2}
\eea
\noindent The expressions for the amplitudes ${\vert A_i \vert}$ and the
relative phase $\theta_1-\theta_2$ are derived from the defining relations
for the Stokes variables. On the other hand, time-evolution of $\theta_1$
is determined from Eq. (\ref{dimer-blg}) by using $A_j={\vert A_j \vert} e^{i
\theta_j}, j=1, 2$. The four integration constants $C_i$ may be chosen
appropriately to implement a variety of initial conditions. There may be
restrictions on the parameters for specified initial conditions such that
${\vert A_1 \vert}, {\vert A_2 \vert}$ are semi-positive definite and phases
are well defined.

The system admits a stationary mode for which $A_1,A_2$ are periodic
in time with constant amplitudes. In particular, the amplitudes
${\vert A_1 \vert}, {\vert A_2 \vert}$ are independent of time, while
the phases depend on time. The stationary solution is obtained by choosing the
integration constants as $C_2=\frac{a}{\eta} C_1, C_3=C_4=0$ for which
$Z^T=\frac{C_1}{\eta} (-c, -b, a)$ and the integration constant $C_1$ may
be fixed through appropriate initial condition. The expressions for $A_1, A_2$
corresponding to this particular solution of $Z$ are obtained by using
Eqs. (\ref{inv-stokes-1}) and (\ref{inv-stokes-2}),
\bea
A_1 & = & \sqrt{\frac{C_1 \beta_0}{\eta}} \ 
e^{i \frac{2 \eta^2 + 3 C_1 \beta_0}{4 \eta} \epsilon^2 t},\nonumber \\ 
A_2 & = & \sqrt{\frac{C_1}{2 \eta} (2 \beta_0+3 \alpha C_1)}
\ \ e^{i \left [ \frac{ 2 \eta^2 + 
3 C_1 \beta_0}{4 \eta} - \tan^{-1} \left ( \frac{2 \Gamma_0}{\eta} \right )
\right ] \epsilon^2 t}.
\eea 
\noindent These solutions are physically acceptable in regions of the
parameter-space determined by Eq.(\ref{real-con}) along with the additional
conditions:
\bea 
\textrm{For} \ \ \alpha \geq 0: C_1 \beta_0 > 0,\ \
\textrm{For} \ \ \alpha < 0:  0 < C_1 \beta_0 < \frac{2 \beta_0^2}{3
{\vert \alpha \vert}}.
\eea
\noindent It may be noted that $C_1$ can always be chosen satisfying these
conditions for any given set of values for $\alpha, \beta_0, \Gamma_0$.
The power $P_i= {\vert A_i \vert}^2$ for the $i^{th}$ wave-guide remains
the same throughout the time-evolution, without being effected by the
loss gain terms. Such a stationary mode, which exists
for ${\cal{PT}}$-symmetric dimer models, is also seen in this
non-${\cal{PT}}$-symmetric Hamiltonian system. Moreover, the allowed
ranges of $\Gamma_0$ can be varied at ease by choosing appropriate
value of the integration constant $C_1$ for a fixed set of parameters
$\alpha$ and $\beta$. This is an advantage over the previous models.

Solutions with time-dependent amplitude as well as phase can also be constructed.
For example, the initial profile $Z(0)^T=(0, 0, 1)$ may
be implemented by choosing $C_1=\frac{2}{3 \alpha \Gamma_0}(1-2 \Gamma_0 \beta_0),
C_2=\frac{a^2}{\eta^2}, C_3=C_4=-\frac{b^2+c^2}{2 \eta^2}$ for which
$Z(t)$ has the following expression:
\bea
Z(t)= \frac{1}{\eta^2}  \begin{pmatrix} 1\\
a b\\ \eta^2-a^2 \end{pmatrix} \cos ( \eta \epsilon^2 t ) -  
\frac{1}{\eta^2}  \begin{pmatrix} 1\\
a b\\-a^2 \end{pmatrix}
+ \frac{1}{\eta} \begin{pmatrix} b \\
-c\\ 0 \end{pmatrix} \sin(\eta \epsilon^2 t).
\eea
\noindent The solution for $A_1, A_2$ may be determined by using the Eqs.
(\ref{inv-stokes-1}) and (\ref{inv-stokes-2}). The Hamiltonian ${\cal{H}}$
is not ${\cal{PT}}$ symmetric, yet it
admits periodic solutions. This ascertains that systems with balanced loss and
gain may admit periodic solutions without any ${\cal{PT}}$ symmetry of the
governing Hamiltonian. The periodic solutions become unbounded for the
values of the parameter for which $\eta$ is imaginary. The corresponding
solutions in terms of hyperbolic functions may be obtained by taking the limit
$\eta \rightarrow i \eta$ in $Z(t)$.

\section{Conclusions $\&$ Discussions}

It has been shown that a non-${\cal{PT}}$ symmetric Hamiltonian system
with balanced loss and gain may admit stable periodic solutions in some
regions of the parameter-space. The result is important from the viewpoint
that all previous investigations are mainly based on ${\cal{PT}}$-symmetric
systems in which the existence of stable periodic solution is attributed to
the unbroken ${\cal{PT}}$-phase. The requirement of ${\cal{PT}}$ symmetry
is too restrictive and there is no compelling reason for a system with balanced
loss and gain to be ${\cal{PT}}$-symmetric in order to admit stable periodic
solutions. The result of this article paves the way for accommodating a large
class of non-${\cal{PT}}$ symmetric Hamiltonian in the mainstream of
investigations on systems with balanced loss and gain. Further, all the
advantages associated with a Hamiltonian system may be used to explore such a
model in detail. 

A coupled Duffing oscillator Hamiltonian system with balanced loss and gain
has been considered as an example to present the results. The Duffing
oscillator is coupled to an anti-damped harmonic oscillator such that the
coupling term effectively acts as a forcing term, albeit in a non-trivial
way. The frequency of the anti-damped oscillator depends on the degree of
freedom corresponding to the Duffing oscillator. There is an interesting
limit in which the dynamics of the Duffing oscillator completely  decouples
from the system, while the anti-damped oscillator is unidirectionally
coupled to it. This limit corresponds to a Hamiltonian formulation for the
standard Duffing oscillator. It should be emphasized that even in this
limit the anti-damped oscillator is not a time-reversed version of the standard
Duffing oscillator. This opens the possibility of investigating the dynamics of
the standard Duffing oscillator using techniques associated with a Hamiltonian
system. Further, the quantum Duffing oscillator may also be introduced and
investigated within the canonical quantization scheme.

It has been shown that the coupled Duffing oscillator model admits stable
periodic solution in some regions of the parameter-space. The Hamiltonian
is non-${\cal{PT}}$-symmetric and there is no question of attributing these
periodic solutions to an unbroken ${\cal{PT}}$-phase. These solutions are
investigated by using perturbative as well as numerical methods. It is known
that the driven Duffing oscillator admits chaotic behaviour. The coupled
Duffing oscillator model investigated in this article also admits chaotic
behaviour in some regions of the parameter-space where the coupling to the
anti-damped oscillator effectively acts as a driving term. This is an
example of a Hamiltonian chaos for systems with balanced loss and gain which
has not been observed earlier.

The method of multiple scale analysis has been used to investigate the
system perturbatively. The amplitude depends on a slower time-scale than
the phase and the dynamics of the amplitude is determined by a set of
coupled nonlinear equations which describe a dimer system. The resulting
dimer model in the leading order of the perturbation for small coupling $\beta$
and loss-gain parameter $\Gamma$ is also Hamiltonian and non-${\cal{PT}}$
symmetric. Further, it is exactly solvable and admits stable periodic solutions
in some regions of the parameters space. This provides an example of
a non-${\cal{PT}}$-symmetric dimer model admitting stable periodic solution.
It should be mentioned here that the dimer model obtained by considering
the nonlinear coupling $\alpha$ as a small parameter is also non-${\cal{PT}}$
symmetric and no exact solutions can be found for the generic values of $\beta$
and $\Gamma$. However, stable periodic solutions are obtained for $\Gamma$
and $\beta$ within a range specified by the linear stability analysis.
It is known that dimer models with balanced loss and gain are important in
the field of optics and provide many counter-intuitive results. The examples
provided in this article suggest that non-${\cal{PT}}$-symmetric systems
should be included within the ambit of the investigations on dimer models
with balanced loss and gain

It is worth recalling some of the results pertaining to ${\cal{PT}}$-symmetric quantum systems
\cite{ben-2,ali,pkg-ph,af,pt-qm-0,pt-qm-1} to place the results obtained in this article
in proper perspective. The general understanding on
non-hermitian quantum system is that it may admit entirely real spectra with
unitary time-evolution provided at least one of the following conditions is satisfied:
\begin{itemize}
\item The Hamiltonian is ${\cal{PT}}$ symmetric and unbroken ${\cal{PT}}$-phase exists\cite{ben-2}.
It may be noted in this context that, unlike in the case of classical mechanics, the time-reversal
symmetry is not unique for quantum system. The non-conventional representation of the time-reversal
operator ${\cal{T}}$ has been used in the literature\cite{pkg-ds,ds-pkg1}.

\item The Hamiltonian $H$ is pseudo-hermitian with respect to a positive-definite similarity operator $\eta$, i.e.
$H^{\dagger} = \eta H \eta^{-1}$\cite{ali}, where $H^{\dagger}$ denotes the hermitian adjoint of $H$.
The system admits an anti-linear symmetry\cite{ali} which may be identified as ${\cal{PT}}$ symmetry
for some special cases. This allows to include non-${\cal{PT}}$-symmetric Hamiltonians with pseudo-hermiticity or
with specific anti-linear symmetry in the main stream of investigations on non-hermitian systems
admitting entirely real spectra and unitary time-evolution\cite{pt-qm-0,pt-qm-1}.
\end{itemize}
\noindent The situation changes significantly for a classical system for which the
time-reversal symmetry is unique and there is no analogue of pesudo-hermiticity or
anti-linear symmetry for the classical Hamiltonian. It appears that the criterion
based on ${\cal{PT}}$-symmetry alone is not sufficient to predict the existence
of periodic solution in a classical balanced loss-gain system. A possible resolution
of the problem  may be to fix the criterion based on the corresponding quantum system
so that anti-linear symmetry and/or pseudo-hermiticity of the quantized Hamiltonian
is used. However, an implementation of the scheme is tricky and nontrivial,
since there may be more than one quantum system for a given classical
Hamiltonian based on the quantization condition. A unique identification of the
quantized Hamiltonian corresponding to a given classical system with balanced
loss and gain that admits periodic solution requires additional
conditions to be imposed. The problem to fix an appropriate criterion
for the existence of periodic solution in classical system with balanced loss and gain remains
unresolved and requires further investigations.

\section{Acknowledgements}

This work of PKG is supported by a grant ({\bf SERB Ref. No. MTR/2018/001036})
from the Science \& Engineering Research Board(SERB), Department of Science
\& Technology, Govt. of India under the {\bf MATRICS} scheme. The work of PR
is supported by CSIR-NET fellowship({\bf CSIR File No.:  09/202(0072)/2017-EMR-I})
of Govt. of India.
\section{Appendix-I: Perturbative solution for $\Gamma \ll 1, \alpha \ll 1$}

Introducing a small parameter $\epsilon \ll 1$ and defining $\Gamma= 
\epsilon \Gamma_0, \alpha=\epsilon \alpha_0$, Eq. (\ref{col-eqn}) can be 
rewritten as,
\bea
\ddot{X} + P X + \epsilon \left [ 2 \Gamma_0 \sigma_3 \dot{X}
+ \alpha_0 \tilde{V}(x) \right ] =0.
\label{col-eqn-1}
\eea
\noindent The unperturbed part of the system is described by
coupled harmonic oscillators satisfying the equation $\ddot{X} + P X=0$.
The terms with the coefficient $\epsilon$ in Eq. (\ref{col-eqn-1}) is
treated as perturbation, which contain the effect of loss-gain and nonlinear
coupling. The standard perturbation theory fails and the method of multiple
time-scales will be employed to analyse Eq. (\ref{col-eqn-1}). The coordinates
are expressed in powers of the small parameter $\epsilon$ and multiple
time-scales are introduced as follow,
\bea
T_n= \epsilon^n t, \ \
X = \sum_{n=0}^{\infty} \epsilon^n X^{(n)}(T_0, T_1, \dots).
\label{var-scale-1}
\eea
\noindent Using Eq. (\ref{var-scale-1}) in Eq. (\ref{col-eqn-1}) and equating
the terms with same coefficient $\epsilon^n$ to zero,
the following equations up to $O(\epsilon)$ are obtained as follows:
\bea
\label{zero.1}
{\cal{O}}(\epsilon^0): && \frac{\partial^2 X^{(0)}}{\partial T_0^2} +
P X^{(0)}=0,\\
{\cal{O}}(\epsilon): && \frac{\partial^2 X^{(1)}}{\partial T_0^2} +
P X^{(1)} + 2 \frac{\partial^2 X^{(0)}}{\partial T_0 \partial T_1} +
2 \Gamma_0 \sigma_3 \frac{\partial X^{(0)}}{\partial T_0} +
\alpha_0 \begin{pmatrix}
x_0^3\\ 3 x_0^2 y_0 \end{pmatrix} =0.
\label{order1.1}
\eea
\noindent These equations are to be solved consistently  to get the
perturbative results.

The unperturbed Eq.  (\ref{zero.1}) has the solution,
\bea
X^{(0)} = A_0 \ e^{-i {\chi_1} T_0}
\begin{pmatrix}
1 \\ \beta
\end{pmatrix}
+ B_0 \ e^{-i {\chi_2} T_0}
\begin{pmatrix}
1 \\ -\beta
\end{pmatrix} + c.c., \ \chi_1=\sqrt{1+\beta}, \ \chi_2=\sqrt{1-\beta}.
\eea
\noindent The $T_1$ dependence of $A_0$ and $B_0$ are determined by the
equations,
\bea
\frac{\partial A_0}{\partial T_1} = - \frac{3 i \alpha_0}{\chi_1}
{\vert A_0 \vert}^2 A_0, \ \
\frac{\partial B_0}{\partial T_1} = - \frac{3 i \alpha_0}{\chi_2}
{\vert B_0 \vert}^2 B_0, \ \
\label{t3-eqn}
\eea
\noindent which have been obtained by eliminating secular terms of Eq.
(\ref{order1.1}). These two equations define a Hamiltonian system,
\bea
{\cal{H}}_1=3 \alpha_0 \left [ \frac{{\vert A_0 \vert}^4}{\chi_1} + \frac{{\vert B_0 \vert}^4}{\chi_2}\right ],
\eea
\noindent with the canonical conjugate pairs as $(A_0, i A_0^*)$ and $(B_0, i B_0^*)$. 
It immediately follows that both ${\vert A_0 \vert}$ and
${\vert B_0 \vert}$ are constants of motion and the constant values are
chosen to be their value at $t=0$. The approximate solution of $X$ is obtained
as,
\bea
X = {\vert A_0(0) \vert}  \ e^{-i t \left [ {\chi_1} + \frac{3 \alpha {\vert
A_0(0) \vert}^2}{\chi_1} \right ]}
\begin{pmatrix}
1 \\ \beta
\end{pmatrix}
+ {\vert B_0(0) \vert} \ e^{-i t \left [ {\chi_2} +
\frac{3 \alpha {\vert B_0(0) \vert}^2}{\chi_2} \right ] }
\begin{pmatrix}
1 \\ -\beta
\end{pmatrix} + c.c. + {\cal{O}}(\epsilon), 
\eea
\noindent which is periodic and has uniform expansion for $t \leq
\epsilon^{-2}$. It may be noted that $\alpha = \alpha_0
\left ( \frac{\Gamma}{\Gamma_0} \right ) $ and the solution inherits the
effect of both the loss-gain and nonlinear interaction.

\end{document}